\documentclass[rmp,nofootinbib,onecolumn,floatfix, twoside,12pt]{revtex4}

\usepackage{graphicx}
\usepackage{amssymb,amsmath,amsfonts,amsthm,epsf,epsfig}
\usepackage{color}
\usepackage[centerlast,small]{caption}
\bibpunct[, ]{[}{]}{,}{a}{,}{,}
\ifnum\lefthyphenmin<3\lefthyphenmin=2\fi
\ifnum\righthyphenmin<2\righthyphenmin=2\fi

\newcommand{\beq}{\begin{equation}}
\newcommand{\eeq}{\end{equation}}
\newcommand{\barr}{\begin{array}}
\newcommand{\earr}{\end{array}}
\newcommand{\ssz}{\scriptsize}

\begin{document}



\title{Spectroscopy of a canonically quantized horizon}
\author{Mohammad H. Ansari}          
\affiliation{ {\ssz University of Waterloo}, {\ssz Perimeter
Institute}}
\thanks{Date: \today} \email{mansari@perimeterinstitute.ca}      
\markboth{\sc \ssz Spectroscopy of a canonically quantized
horizon}{\sc \ssz M. Ansari}

\begin{abstract}
  \centering
\begin{tabular}{p{11cm}}
\setlength{\baselineskip}{.6\baselineskip} \textbf{Summary:}
Deviations from Hawking's thermal black hole spectrum, observable
for macroscopic black holes, are derived from a model of a quantum
horizon in loop quantum gravity. These arise from additional area
eigenstates present in quantum surfaces excluded by the classical
isolated horizon boundary conditions. The complete spectrum of area
unexpectedly exhibits evenly spaced symmetry. This leads to an
enhancement of some spectral lines on top of the thermal spectrum.
This can imprint characteristic features into the spectra of black
hole systems. It most notably gives the signature of quantum gravity
observability in radiation from primordial black holes, and makes it
possible to test loop quantum gravity with black holes well above
Planck scale.
\end{tabular}
\end{abstract}
\maketitle

%

{ \setlength{\parskip}{-1ex }\tableofcontents}

\setlength{\baselineskip}{.8\baselineskip}

\setlength{\parskip}{1ex plus 0.5ex minus 0.2ex}


\section{Introduction}

Most astrophysicists agree that black holes exist and radiate. So
far three types of black hole radiations have been investigated:
($i$) the Hawking radiation, ($ii$) the gravitational radiation, and
($iii$) the X-ray emission from the infalling materials into a black
hole. In this note, the quantum geometry of the horizon is, under
certain assumption, shown to imply revision of the first type of
black hole radiation.

The Hawking radiation is known semi-classically to be continuous.
However, the Hawking quanta of energy are not able to hover at a
fixed distance from the horizon since the geometry of the horizon
has to fluctuate, once quantum gravitational effects are included.
Thus, one suspects a modification of the radiation when quantum
geometrical effects are properly taken into account. Any transition
between two horizon area states can affect the radiation pattern of
the black hole. The quantum fluctuations of horizon may either
modify, alter or even obviate the semiclassical spectrum,
\cite{{Helfer:2003va},{Wald:1999vt}}.

Bekenstein and Mukhanov in \cite{Bekenstein:1995ju} studied a simple
model of the quantum gravity of the horizon in which area is equally
spaced. They found no continuous thermal spectrum but instead black
holes radiate into discrete frequencies.  The natural width of the
spectrum lines turns out to be smaller than the energy gap between
two consecutive lines. Thus, their simple model predicts a
falsifiable discrete pattern of {\it equidistant} lines which are
unblended. This result is not completely in contradiction with
Hawking prediction of an effectively continuous thermal spectrum of
black hole using semiclassical method, since the discrete line
intensities are enveloped in Hawking radiation intensity pattern.

More recently, it has been possible to study the quantum geometry of
horizons using precise method in loop quantum gravity. In this
non-perturbative canonical approach, the quantum geometry is
determined by geometrical observable operators. Canonical
quantization of geometry supports the discreteness of quantum
area.\footnote{A summary of emergent aspects of non-stringy quantum
gravity theories can be found in \cite{Smolin:2006pa},
\cite{Markopoulou:2006qh}, and the references therein.} This theory
does not reproduce equally spaced area, instead the quanta become
denser in larger values, \cite{{Smolin:1994ge},{Ashtekar-surface}}.
Having defined a black hole horizon as an internal {\it boundary} of
space \cite{Ashtekar:black hole}, only a subset of area eigenvalues
contribute to identifying the horizon area. In fact, this subset
contains the area associated to the edges puncturing the boundary.
This subset is not evenly spaced and it turns out that the area
fluctuations of such a horizon do not imprint quantum gravitational
characteristics on black hole radiation, \cite{Barreira:1996dt}.

\begin{figure}
  \includegraphics[width=6.5cm]{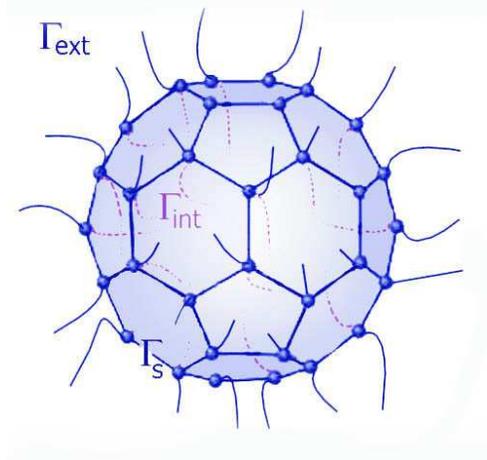}\\
  \caption{A quantized black hole}\label{fig. black hole}
\end{figure}

Nonetheless, restricting the quanta of horizon area to the subset of
punctures is based on a non-trivial gauge-fixing of the horizon
degrees of freedom. This is sufficient for the purpose of black hole
entropy calculation since it results to the residence of a finite
number of degrees of freedom on the horizon, independently from the
bulk.  Such a quantization, while is too restrictive, leaves some
physical ambiguities. For instance, in classical general relativity
spacetime metric field does not {\it end} at a black hole horizon,
instead it extends through the black hole. In fact, a quantum black
hole in a space manifold, instead of being the reason for
termination of quantum space, partitions it into three subgraphs: 1)
the partition that reside outside of horizon,
$\Gamma_{\mathrm{ext}}$, 2) the partition that reside inside of the
horizon, $\Gamma_{\mathrm{int}}$, and 3) the partition that lies on
the horizon 2-surface, $\Gamma_{\mathrm{s}}$. On the horizon surface
some vertices and completely tangential edges reside. The spin
network states associated to a partition that consists of the
vertices lying on the horizon are called {\it horizon spin network
states}. These states determine the whole quantum geometry of the
underlying horizon, Fig (\ref{fig. black hole}). Under some
simplifications, the spin network state associated to a spherical
symmetric structure has been worked out in \cite{Bojowald:2004si}.
The quanta of such a horizon area is chosen from the complete
spectrum. It reproduces the Bekenstein-Hawking entropy
\cite{Ansari:2006cx}. Moreover, in this note it is shown that such a
black hole exhibits unexpectedly a macroscopic effect in the black
hole radiation.

The aim of this note is two fold:
\begin{enumerate}
  \item Firstly, in Part (\ref{part1}) an unexpected symmetry in the complete spectrum
of area is descried. In fact, this spectrum can be decomposed into a
several {\it evenly} spaced sets, each with individual gap between
levels. This leads to a reduced formula of area eigenvalues. In
$SU(2)$ version of loop quantum gravity the gaps scale as the square
roots of `square-free' numbers. In $SO(3)$ version, they are the
square roots of the discriminants of all possible quadratic positive
definite forms.

  \item Secondly, in Part (II) it is discussed that having applied the
complete spectrum of area, a black hole radiates quantum
mechanically a continuous spectrum. But the existence of the
symmetry within the area spectrum results to a phenomenon called the
{\it quantum amplification effect}. This generates several distinct
bright lines in radiance spectrum. It gives the signature of quantum
gravity observability in radiation from primordial black holes.
Moreover, it challenges the isolated horizon picture conjecture,
while makes it possible to test loop quantum gravity with black hole
radiation well above Planck scale.
\end{enumerate}

Before these, some of the attempts to discovering the signature of
quantum gravity in a black hole radiation are reviewed.


\section{Some theories} \label{sec Review}

Firstly, a model of quantum gravity that predicts macroscopic
effects on black hole radiation is reviewed. Afterwards the attempts
within loop quantum gravity are illustrated.

\subsection{A quantum geometry and black hole radiation}
A sector of spacetime may collapse and settles down to a stationary
state in which the zeroth law of black hole mechanics is satisfied;
the surface gravity is constant over the event horizon of the
sector. The sector is called black hole. The ADM mass of a neutral
non-rotating black hole non-trivially depends on the black hole
horizon area. This is the first thermodynamic law of black holes,

\begin{equation}\label{eq. A=M^2}
    A = \frac{16 \pi G^2}{ c^4} M^2.
\end{equation}

Steven Hawking uncovered that quantum field theory in black hole
curved spacetime leads to particle creation effect at the horizon,
thus black hole radiates. The original derivations of this radiation
was made of particle propagating into the black hole, the radiation
is independent of the notion of particle, \cite{Fredenhagen:1989kr}.
The sum of the black hole entropy plus the matter entropy outside
the black hole never decreases, $S_{\mathrm{outside}} +
S_{\mathrm{black\ Hole}} \geq 0$. This is the generalized second
thermodynamic law of black hole. This law holds even during quantum
evaporation of the black hole via Hawking radiation, when a negative
energy flux across the horizon decreases of area. Although, the way
a black hole loses mass during the thermal radiation implicitly must
involve quantum gravitational assumptions.

Jacob Bekenstein and Venceslav Mukhanov postulated a rough theory of
quantum gravity in which the horizon area of a black hole is
quantized in uniformly spaced tiny fractions of the Planck length
scale,

\begin{equation}\label{eq. Bek ansatz}
A =  \alpha  n\ \ell^2_{\mathrm{P}},
\end{equation}
where $n$ is a natural number, and $\ell_{\mathrm{P}}$ is Planck
length, $\sqrt{\hbar G/ c^3} \sim 1.6\times 10^{-35}\mathrm{m}$,
which is drastically small.

Semiclassically, the discreteness of the quantum values of a horizon
area leads to the discreteness of  black hole mass. If a black hole
is defined as a quantum system in thermodynamical equilibrium, the
radiation is analogous to quantum mechanical instability that leads
to quantum decays. Having the energy levels of a non-rotating
neutral black hole, the transitions between neighboring energy
levels causes quantum decay. A discrete mass spectrum implies the
discreteness of mass emissions. From (\ref{eq. Bek ansatz}) and
(\ref{eq. A=M^2}) the quanta of energy are

\begin{equation}\label{eq. deltaM=}
   \delta M = \frac{\alpha \delta n}{32\pi M} M_{\mathrm{P}},
\end{equation}
 where $M_{\mathrm{P}}$ is the Planck mass, $\sqrt{\hbar c/G}\sim 2.2
\times 10^{-8}$ kg.

Under the assumption that the black hole mass is not changed during
the quantum emissions, $\delta M \ll M$, and by the use of (\ref{eq.
deltaM=}), the frequencies of emissive quanta turn out to be
integers multiplied by a minimal frequency. The minimum frequency is
called {\it the fundamental frequency} $\varpi$,
\begin{equation}\label{eq. Bek w fund}
  \varpi = \frac{\alpha c^3}{32 \pi G M}.
\end{equation}

Other emissive frequencies are  {\it harmonics} $\omega_n$, which
are proportional to this fundamental frequency by an integer $n$,
$\omega_n=n\varpi$.

Under the assumption of uniform matrix elements of quantum
transitions between near levels, the intensities of the spectral
lines were worked out in \cite{Bekenstein:1995ju}. The outcome turn
out to be enveloped by the Hawking radiation intensity, whilst the
allowed frequencies are discrete and equidistant. Moreover, it turns
out that the thermal broadening of the lines are smaller than the
gap between any two consecutive harmonics.

From the model three major conclusion come about, ($i$) there should
be no lines with wavelength of the order of the black hole size or
larger, ($ii$) the black hole radiance spectrum must be clearly
discrete and the lines do not overlap, ($iii$) the radiance pattern
is a  {\it uniformly spaced} discrete lines.

Nonetheless, there has not been any justification for this evenly
spaced area from within the very quantum gravitational theories. In
the next sections we consider a version of quantum gravity whose
roots are within the so-called loop quantum gravity.


\subsection{A quantum geometry and isolated horizon radiation}

The first suggestion to describe a black hole as a 2-surface
boundary of space manifold in loop quantum gravity was proposed by
Krill Krasnov in \cite{Krasnov:1996tb}. Carlo Rovelli based on the
picture discussed the black hole entropy in \cite{Rovelli:1996dv}.
Afterwards, by the developments of the isolated horizon theory the
bounded sector was more precisely defined in a series of works,
\cite{Ashtekar:black hole}.

A black hole is a classical concept and its definition is highly
non-local, because one has to know the information of the entire
spacetime manifold $(\Sigma, g_{\mu\nu})$, in order to find the
entire causal past of the future null infinity.  A black hole is a
sector of manifold that does not intersect with the entire past of
the future null infinity. However, this definition is not
well-suited for the purpose of identifying a black hole region in a
non-perturbative canonical quantized space. In fact, a more local
criteria must be installed on such a theory.

A classical isolated horizon is defined by a set of boundary
conditions of a sector of spacetime $\Delta$, which mimics the
essential local structure of a static event horizon. Assuming the
black hole sector to be $S^2\times \mathbb{R}$, these boundary
conditions are necessary to verify the black hole thermodynamic laws
from the sector: 1) the Einstein equations hold at the sector,
  2) the sector is null,
  3) the sector is equipped with a preferred foliation
  by 2-spheres transverse to its null normal $l^a$; the second null normal to $S_{\Delta}$ is $n^a$ with $l^a n_a =
  -1$,
  4) the sector is non-rotating,
  5) $l^a$ is twist-, shear-, and expansion-free geodesic; $n^a$ is twist- and shear-free with negative expansion
  $\theta_{(n)}$,
  6) $\theta_{(n)}$ is constant over each foliated shell
  $S_{\Delta}$,
  7) the flux densities of electric and magnetic fields are
  uniform through each $S_{\Delta}$.

\begin{figure}[h]
 \includegraphics[width=4cm]{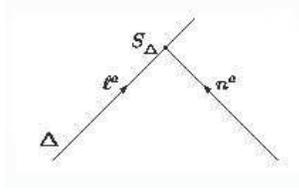}
  \caption{The black hole sector $\Delta$, its two null normals and its preferred foliation $S_{\Delta}$.}
  \label{fig. delta}
\end{figure}

On the other hand, it is known that general relativity can be
written in terms of gauge filed. For this aim, a trivial $SU(2)$
bundle is assumed over the space 3-manifold. For each positive real
number $\gamma$, a phase space $ ^{\gamma}\Pi$ is assumed to exist.
This phase space consists of the configuration fields, the
connection fields $ ^{\gamma}A^i_a$ (1-form), and the canonical
momenta, the fields $^{\gamma} \tilde{\Sigma}_{ab\ i}$ of density
weight one (2-form). $i=1,2,3$ the gauge degrees of freedom and
$a=1,2,3$ the spatial degrees of freedom. The curvature of the
connection field is a 2-form field $^{\gamma}F^i_{ab}$. The Einstein
equations for any $\gamma$ is verified.

The so-called `triad fields' $ ^{\gamma} E^a_i$ are defined via the
momentum fields $\tilde{E}^a_i := \gamma \varepsilon^{abc} \
^{\gamma} \tilde{\Sigma}_{bc\ i}$, where $\varepsilon^{abc}$ is
Levi-Civita $\varepsilon^{abc}$ of density weight one. From the
triads the 3-metric variables $q^{ab}$ are defined,
$^{\gamma}\tilde{E}_i^a\ ^{\gamma}\tilde{E}^{bi} = q q^{ab}$. Also,
the triad fields define area of a 2-surface. Since the area of a
2-surface is $\int_{S} \sqrt{q}d^2x$, given the relation between the
momenta and the 3-metric, the area can be redefined as $\int_S d^2x
\sqrt{ ^{\gamma}E^a_i n_a\ ^{\gamma}E^b_i n_b}$, a functional of
momenta, where $n_a$ is the normal to the surface.

In this language, a neutral stationary black hole in the manifold is
the problem of adding a boundary with special boundary condition to
the theory. The black hole sector is $\Delta$ where it is foliated
by $S^2 \times \mathbb{R}$. The boundary of the sector,
$S_{\Delta}$, must satisfy the above mentioned conditions of an
isolated horizon. In the gauge language of gravity, there is a way
to define two null vectors of desired properties $l^1n_a=-1$, $l^a
l_a=n^an_a=0$ by the use of triad momentum conjugate fields
$^{\gamma} \tilde{\Sigma}_{ab\ i}$. Having the two null vectors the
following conditions must be imposed further in order to make a
quasi-local black hole:

\begin{description}
  \item[Area-fixing:] the manifold momenta
  must admit a {\it fixed} value of area $a$ on the shell.

  \item[Gauge-fixing:] the pullback
  $\overleftarrow{^{\gamma}A^i_a}$  of the bulk
  connection fields to the shell
  $S_{\Delta}$ are the $U(1)$ connection
  fields $^{\gamma}W_a$,  up to a constant. For this aim, a $U(1)$ sub-bundle is selected at the shell
$S_{\Delta}$. By fixing a unit vector $r^i$ at every point of the
shell, the connection field on the sub-bundle will be $u(1)$-valued
$^{\gamma}W_a$.

  \item[Boundary condition:] the pullback
  $\overleftarrow{^{\gamma}\Sigma_{ab\ i}}$ of the
  bulk momenta to the shell
  $S_{\Delta}$  are completely determined by the curvature $^{\gamma}F_{ab} =
\partial_a\ ^{\gamma}W_b -
\partial_b\ ^{\gamma}W_a $. The relation between these two turns
to be $^{\gamma}F_{ab} = -(2 \pi \gamma/a)\
\overleftarrow{^{\gamma}\Sigma_{ab\ i}} r^i$.

   \item[The field equations:] at the sector the equations
   of motion hold.
\end{description}

 The contribution of the boundary in the gravitational action is the addition of a
 $U(1)$ Chern-Simons action term of the gauge fields $^{\gamma}W_a
 $. Such an action is invariant under the following
transformations: 1) $SU(2)$ gauge transformation, those reduce to
$U(1)$ transformation on $S_{\Delta}$ and identity on the infinity,
2) spatial diffeomorphism, those reduce to tangent transformation on
the shell and the identity at the infinity, 3) time evolution
between the fixed horizon and infinity with lapse going to zero at
the horizon and a constant at the infinity, 4) phase space
transformation between different $\gamma$-sectors $^{\gamma}\Pi$ and
$^{\gamma'}\Pi$.\footnote{In the quantum version, the quantum phase
space $ ^{\gamma}\Pi$ is unitarily {\it inequivalent} to the one of
another quantum phase space $ ^{\gamma'}\Pi$.}

Such a classical horizon does not carry independent degrees of
freedom due to the existence of the strong boundary condition.
However, the quantum version is different.

To quantize the theory, a graph is embedded into the manifold and
the connection fields are generalized into $su(2)$-valued holonomies
along the pathes of the graph. Two Hilbert spaces are obtained, the
one of the bulk $\mathcal{H}_{\mathrm{bulk}}$ and the one of the
boundary $\mathcal{H}_\mathrm{boundary}$. The boundary Hilbert space
is defined on the Chern-Simons charged points, namely `punctures' of
the surface. The bulk Hilbert space has a basis by {\it spin
networks} in the spatial 3-manifold with `loose ends' at the charge
points of the internal boundary. \footnote{Quantization of a black
hole has not been understood yet. There are several models for this
purpose. Among them those are acceptable that do not make serious
contradictions with the certain classical properties of a horizon.
Let us consider non-perturbative context of quantum gravity. One of
the recent model introduces a quantum black hole based on the action
of `expansion operator' on a the quantum state of a mixture of
geometry and matter, \cite{Husain:2005jx}. There is another model
based on causal dynamical triangulation. The causal dynamical
triangulation is a non-perturbative quantum gravity analytically
worked out in two dimensions in \cite{amb98}, statistically in
\cite{Ansari:2005uz}, as well as numerically in higher dimensions. A
black hole could be defined in this model, \cite{Dittrich:2005sy}.}

Consider a spin network state in the bulk Hilbert space. In this
wave function, the edges of the spin network are labeled by the
irreducible representation of the holonomies (the so-called
`spins'), the vertices are intertwiners, the punctures by a vector
$|m\rangle$ in the representation of the incident edge. If the spin
of the incident edge to the puncture is $j$, there exist $2j+1$
different copies of puncture states, $m \in \{ -j, -j+1, \cdots,
j-1, j \}$.

Notice that each puncture is a place where and edge ends at the
surface and thus it carries the area eigenvalue corresponding to the
edge.

The surface Hilbert space $\mathcal{H}_{\mathrm{boundary}}$ contains
$u(1)$-valued connection fields. The geometry of the surface is flat
except at the punctures, where there are conical singularities. All
of different horizon wave functions corresponding to one edge of
spin $j$ produce the same horizon area, because the area eigenvalues
only depend on $j$. Carlo Rovelli and Lee Smolin verified this area
first by the use of loop operators in \cite{Smolin:1994ge}. They
found that the spectrum of area associated to an edge is `almost'
equidistant in large scales. The area of a puncture depends on the
irreducible dimension of the puncturing edge. In fact, these
eigenvalues of area were those were discovered first. An edge of
spin $j$ generates the area $a_j = 8 \pi \gamma \ell^2_{\mathrm{P}}
\sqrt{j (j+1)}$ on the boundary. These eigenvalues depend only on
one quantum number, $j$.

Having $2j+1$ different copies assigned to the same area, the
horizon wave functions are degenerate, thus black hole gets non-zero
entropy. The entropy is proportional to the surface area and since
the surface is assumed to be of fixed area, the entropy of an
isolated horizon meet the second thermodynamic law of black hole, it
is non-decreasing. Therefore, the entropy is physical.

Later on Abhay Ashtekar and Jerzy Lewandowski derived the complete
spectrum of area operator in \cite{Ashtekar-surface}. The spectrum
that Rovelli and Smolin have discovered was a subset of the complete
spectrum of area.  The complete spectrum is also discrete, although
the eigenvalues approaches to a continuum in large eigenvalues. This
spectrum is described in section (\ref{sec_Horizon_as_a_partition}).

The gravitational fields about a black hole are not stable because
they interact with `non-stationary' matter fields. Only about such a
shell, from the Einstein equation the decreasing of mass by $\Delta
E$ corresponds to the decreasing of area by
 $\Delta A$ such that $\Delta A = 32 \pi G^2 E \Delta E$.
This correlation describes the transitions between two
macroscopically stable states after mass perturbation. A quantum
jumping down an area level corresponds to emission of one (or some)
quantum of area. In both $SU(2)$ and $SO(3)$ versions of loop
quantum gravity small values of spin $j$ produces the quantum of
area proportional to a number within the interval $[j, j+1]$. At
large $j$ the area make it approximately proportional to $j$.
Therefore, a transitions from a high level into a low level does not
coincide with the transitions from a higher level into that high
level. In other words, one-punctural decays produces an effectively
continuous radiance spectrum at high frequencies.

However, the relation (\ref{eq. deltaM=}) since is a classical
relation, does not guarantee the occurrence of only one-punctural
transmissions.  A quantum black hole may also radiate a
multi-punctural decay in its low damped quasinormal modes. A
multi-punctural decay is an emission in which a set of punctures
simultaneously undergo area shrinking in one go and produce one
quantum of energy.  For instance, consider a black hole made of
three patches of area, two of which correspond to punctures of spin
1/2 and the third one to a spin 1. The overall horizon area is $A_1=
8 \pi \gamma (\sqrt{2}+\sqrt{3}) \ell_P^2$. This black hole may
decay into a geometrical configuration with two punctures of spin 1.
In this case the horizon area is shrunk into $A_2=8\pi \gamma
(2\sqrt{2})\ell_P^2$. According to (\ref{eq. deltaM=}) the emitted
energy is proportional to $8\pi \gamma (\sqrt{3}- \sqrt{2})\ell_P^2$
by a constant, which is even smaller than the minimal
single-punctural decay ($8 \pi \gamma \sqrt{2} \ell_P^2$). Such a
typical multi-punctural emissions can take almost any value and fill
the continuous spectrum in all ranges of energy.

Since the puncture quantum of area is not uniformly spaced, the area
fluctuations produces a continuous spectrum o emissive frequencies.
While such a prediction satisfies the Hawking pattern of radiation,
since the populations of all frequencies are uniform, there is no
notable quantization effect in the black hole radiation,
\cite{Barreira:1996dt}. In the next section, a different picture of
black hole is reviewed and its radiance pattern is
analyzed.\footnote{Beside these two possible pictures of a horizon
in loop quantum geometry, there exists also a third one that was
proposed by Livine and Terno in \cite{Livine:2005mw}.}


\subsection{A quantum geometry and spin network horizon radiation}\label{sec_Horizon_as_a_partition}

In this section a new picture of a black hole is explained and the
quantum effects on its radiation is described in the rest of the
note.

In brief, deriving the entropy of an isolated horizon depends on
fixing a gauge of the connections fields. More precisely, in the
presence of such a classical boundary-like horizon in the underlying
manifold, the $su(2)$-valued connection fields of the bulk are
gauge-fixed into $u(1)$-valued connections on the boundary and thus
the punctures take additional degree of freedom independent from
those of the bulk. This assumption is too restrictive. In such a
quantum surface many quanta of horizon area are excluded by the
classical isolated horizon boundary conditions. However, considering
the complete spectrum of area eigenvalues as the possible horizon
area, provides the same entropy that is expected on black hole
horizon, while it gives a different picture of a black hole, a more
quantum picture.

There have been some attempts to define a black hole as a partition
of spin network state. For instance, Martin Bojowald in
\cite{Bojowald:2004si} tried to see a spherical symmetrical black
hole state as a spin network state. In this picture, a black hole
horizon is defined by studying the properties of an infalling spin
network states associated to a 2-surface through the black hole. In
this picture, instead of considering the evolution of the underlying
surface and quantizing the manifold afterwards, the quantum state
evolves itself independently. In fact, the behavior of the quantum
surface in time can narrow the definition of appropriate dynamics of
black hole.

\begin{figure}
  \includegraphics[width=5cm]{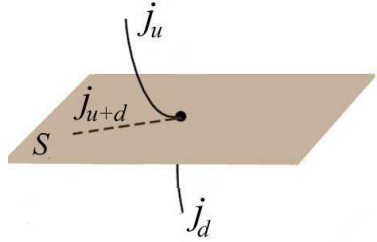}\\
  \caption{Two incident edges at a vertex residing on $S$.}\label{fig. art}
\end{figure}

Let us consider a quantum surface associated to a surface $S$ in a
3-manifold $\Sigma$.  This surface divides the manifold into two
disjoint open sets $\Sigma_{\mathrm{up}}$ and
$\Sigma_{\mathrm{down}}$ such that $\Sigma = \Sigma_{\mathrm{up}}
\cup S \cup \Sigma_{\mathrm{down}}$ with $\Sigma_{\mathrm{up}} \cap
\Sigma_{\mathrm{down}} = \emptyset$, Figure (\ref{fig. art}). Thus,
the imbedding graph $\Gamma$ in the presence of underlying 2-surface
$S$ is split into three subgraphs: ($i$) $\Gamma_{\mathrm{up}}$,
which is completely in one side of $S$ in the 3-manifold, ($ii$)
$\Gamma_{\mathrm{down}}$, which is completely in the other side of
$S$, and ($iii$) $\Gamma_{\mathrm{s}}$, which lies on $S$.
$\Gamma_{\mathrm{s}}$ consists of some residing vertices
$\{v^{\alpha}\}$ on $S$ as well as some tangential edge lying
entirely on the horizon surface $S$.

Consider a typical spin network state corresponding to a residing
vertex on an underlying surface $S$, i.e. the one in figure
(\ref{fig. art}). This state intertwines the bulk edges of external
and internal sub-graphs, and the edges of $\Gamma_{\mathrm{s}}$. The
set of all such spin networks produces a partition of spin network
states called the {\it quantum surface states}. This state is
isolated within the near-surface region. The quantum geometrical
state of the surface is determined by these spin network states. A
bulk edge relative to $S$ falls into three categories: either (a) it
bends tangentially at the point at which the edge crosses the
surface $S$, or (b) it intersects the surface at a point without
bending at the surface, or (c) it lies completely tangential to the
surface. The edges which are completely tangential to $S$, the
so-called `analytical edges', do not belong to the bulk edges,
instead they belong to the quantum surface $S$.

In both the isolated horizon picture and the black hole spin network
a quantum state is associated to the horizon. But in the latter one
horizon is defined classically not quantum mechanically. The
continuum surface undergoes evolution and a static Hilbert space is
associated to the classically evolving surfaces at each time frame.
In other words, the quantum state of horizon {\it follows} what the
underlying surface {\it rules}. The quantization procedure of
gauge-fixing prior to the quantization is not a trivial method of
quantization. In fact, in the case of quantum spacetime such a
quantization causes some ambiguities:
\begin{itemize}
  \item The black hole sector is identified classically
  and remains exactly the same after quantization, without considering any uncertainty on black hole
  radius and its intrinsic geometry.
  \item The spin networks end at the black hole horizon, which contradicts with the classical definition of black hole. In classical general relativity, the metric fields extend through the horizon.
  \item No tunneling effect is allowed throughout the horizon.
\end{itemize}

To overcome the problems, one can treat the quantum horizon as an
evolving quantum surface which undergoes its quantum evolution. The
evolution is only expected to verify the classical results only at
the classical limits. Such a quantum black hole is a partition of
spin network, whose boundary determines its quantum horizon.
However, it is not so easy to define a surface without reference to
a background metric or other fields. One surface that {\it can} be
defined in a background-independent manner is a black hole horizon.
This is a property that distinguishes horizons from most other
surfaces. The final state of this partition should not be influenced
by the initial states of the rest of the rest of the world. The
initial state of this partition should influence the final states of
the rest of the rest of the world. moreover, the entropy associated
to the vertices residing on the horizon remains fixed. Also it is
expected that the quantum sector gets non-expanding volume, as well
as horizon area. This make it possible to make this definition of
black hole more realistic because it the black hole be less hidden
from a quantum system closer to the horizon,
\cite{Ansari2006locality}.

What is the entropy of such a quantum black hole? Considering a
typical spin network state like the one of the figure (\ref{fig.
art}), the action of area operator on this state generates an area
eigenvalue. It turns out that an area eigenvalue corresponds to a
{\it finite} number of different eigenstates. In fact,
$su(2)$-valued quantum states of a surface of certain area are
degenerate state. The degeneracy is such that the entropy of the
surface is proportional to the surface area and in the case of black
holes quantum surface the entropy verifies the Bekenstein-Hawking
entropy, \cite{Ansari:2006cx}. This entropy is not necessarily
non-decreasing in the course of time, unless it is fixed by the
defining the appropriate evolution of horizon quantum surface
states, \cite{Sorkin:2005qx}.\footnote{The question why a horizon
carries physical entropy whilst a random surface does not, is subtle
and still not understood fully. This is not only a property of
canonical quantization of spacetime. For instance, in the causal
dynamical triangulation, which is another non-perturbative approach
to quantum gravity \cite{Ambjorn:2001cv}, for the purpose of
obtaining a 1+1 global geometry (a 2-surface) by triangular building
blocks, the two components of the blocks can be respected as up and
down spins with respect to an external time field,
\cite{Ansari:2005uz}. If one coarse-grain a `spin' by ignoring the
interior of some randomly selected region of the surface, one will
obtain an entropy-like number. However, this number will not have
the properties we normally associate with entropy.}

The complete eigenvalues of area operator on a typical spin network
state was first studied in \cite{{Ashtekar-surface}} and a few
months later the results were verified by the use of recoupling
theory in \cite{{Frittelli:1996cj}}. In fact, the area of a spin
network is the outcome of linking the two sides of the surface. Let
the up and down edges of the vertex $\alpha$ get the spin
$j_u^{(\alpha)}$ and $j_d^{(\alpha)}$, respectively. The two edges
may bend tangentially at the underlying 2-surface at their joint
intersecting vertex. The overall tangent vector induced from them on
the surface take the spin $j_{u+d}^{(\alpha)}$. The spin $j_{u+d}$
take discrete values and bounded to the following values
\begin{equation}
\label{eq. jud} j_{u+d}^{(\alpha)} \in
\{j_u^{(\alpha)}+j_d^{(\alpha)}, j_u^{(\alpha)}+j_d^{(\alpha)} -1 ,
\cdots, |j_u^{(\alpha)} - j_d^{(\alpha)}| +1 , |j_u^{(\alpha)}
-j_d^{(\alpha)}| \}.
\end{equation}

The action of area operator on a typical area state corresponding to
incident edges at a residing vertex on $S$ entangles the external
and internal edge. Let us for simplicity define the color numbers
corresponding to the three spins, $p:=2j_u^{(\alpha)}$,
$q:=2j_d^{(\alpha)}$, and $r:=2j_{u+d}^{(\alpha)}$. The area squared
operator acting on the trivalent state $\langle p, q, r|$ entangles
two sides of the underlying surface (the shaded and unshaded sides),
\begin{eqnarray}
\label{eq. <tri| A = ...} \barr{cccc} \barr{c}
\mbox{\includegraphics[height=1.5cm]{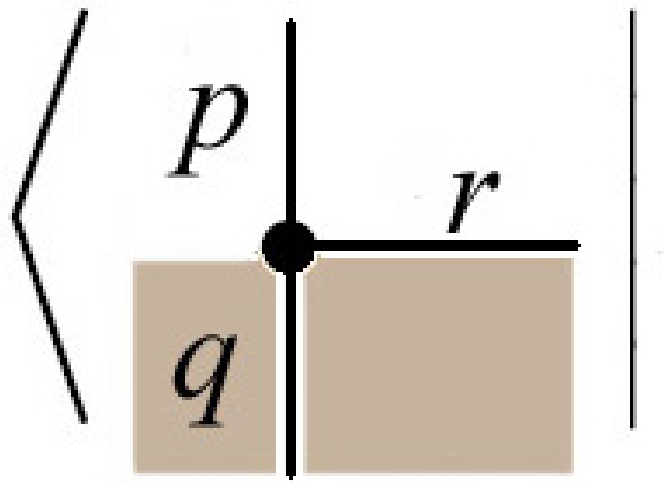}} \earr
 \hat{A}^2 = -b^2 & \left(  p^2
 \barr{c}\includegraphics[width=1.5cm]{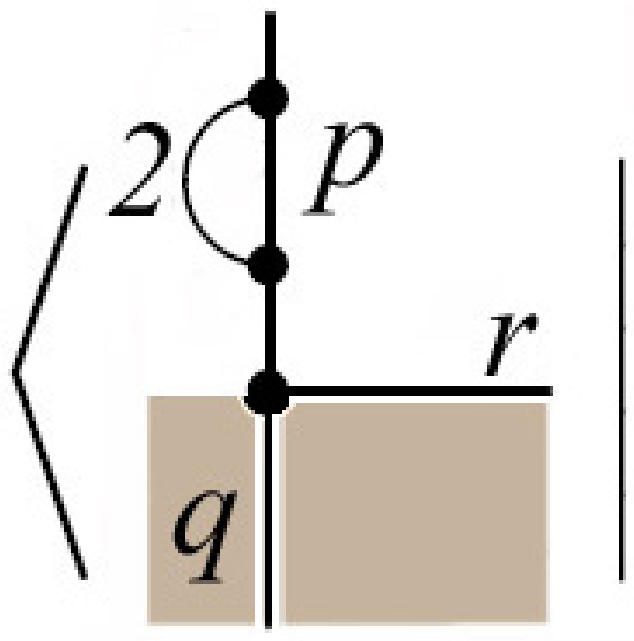} \earr \right.  & + & q^2 \barr{c}\mbox{\includegraphics[height=1.5cm]{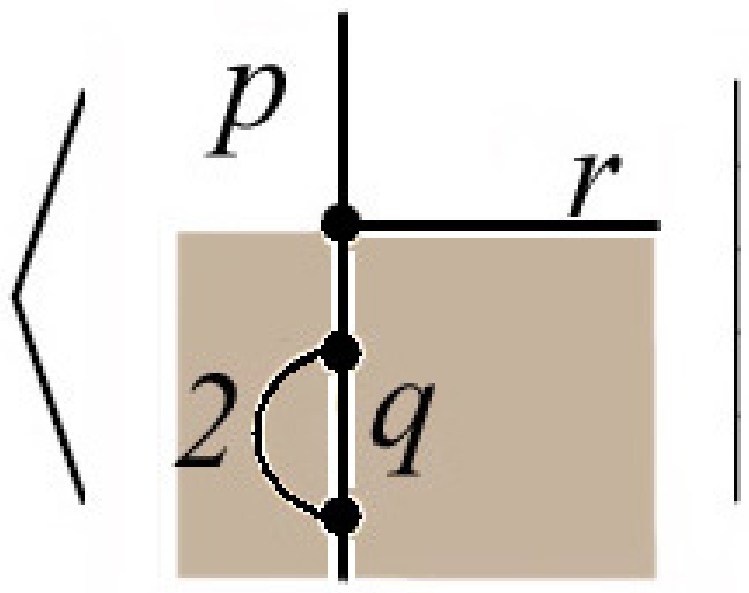}}\earr  \\
&  & + & \left. 2pq\ \
\barr{c}\mbox{\includegraphics[height=1.5cm]{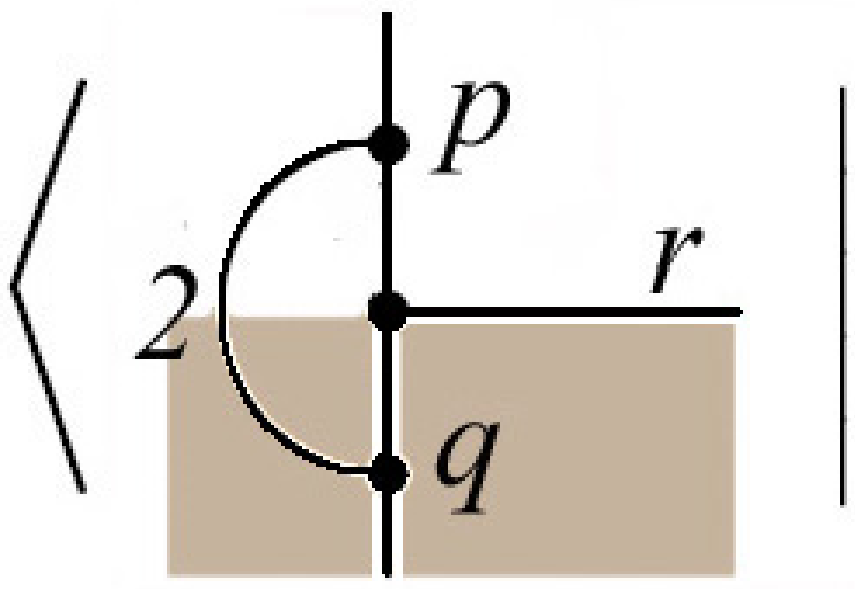}}\earr
\right)\\
\earr
\end{eqnarray}
where $b :=  8 \pi \gamma \ell_P^2$. Using the reduction formulae of
recoupling theory, the grasped states are identical to the original
state,
\begin{eqnarray}
\label{eq. 3 coupling formulae}
\begin{array}{ccc}
\barr{c} \mbox{\includegraphics[height=1.5cm]{p2qr.eps}} \earr &=&
-\frac{(p+2)}{2p} \barr{c}
\mbox{\includegraphics[height=1.5cm]{pqr.eps}} \earr , \\
\barr{c} \mbox{\includegraphics[height=1.5cm]{2pqr.eps}} \earr &=&
\frac{-2p(p+2) - 2q(q+2) + 2r(r+2)}{8pq} \barr{c}
\mbox{\includegraphics[height=1.5cm]{pqr.eps}} \earr .
\end{array}
\end{eqnarray}

Substituting (\ref{eq. 3 coupling formulae}) in (\ref{eq. <tri| A =
...}) the squared area operator acting on $\langle p, q, r|$ turns
out to become an eigenstate relation non-trivially. Thus, the
trivalent area state $\langle p, q, r|$ is the eigenstate of the
operator. Finally, the area eigenvalues corresponding to the cell
$\alpha$ turns out to be $a^{(\alpha)} = (4\pi \gamma)\ m_{j_u,
j_d,j_{u+d}}^{(\alpha)} \ \ell^2_{\mathrm{P}}$, where
\begin{equation}
\label{eq. A degenerate sec.} m_{j_u, j_d,j_{u+d}}^{(\alpha)} =
\sqrt{2j_u^{(\alpha)}(j_u^{(\alpha)}+1) +
2j_d^{(\alpha)}(j_d^{(\alpha)}+1) -
j_{u+d}^{(\alpha)}(j_{u+d}^{(\alpha)}+1)}.
\end{equation}

A  Schwarzschild black hole horizon belongs to the class of surfaces
that has no boundary, $\partial S = \emptyset$ and divide the
3-manifold $\Sigma$ into two disjoint sets
$\Sigma_{\mathrm{internal}}$ and $\Sigma_{\mathrm{external}}$ such
that $\Sigma = \Sigma_{\mathrm{internal}} \cup S \cup
\Sigma_{\mathrm{external}}$ with $\Sigma_{\mathrm{internal}} \cap
\Sigma_{\mathrm{external}} = \emptyset$. Thus, the graph $\Gamma$ in
the presence of a black hole is split into three graphs
$\Gamma_{\mathrm{external}}$, $\Gamma_{\mathrm{internal}}$, and
$\Gamma_{\mathrm{s}}$. Notice that the corresponding states to a
compact closed surface can only gauge transform into another compact
closed state. Therefore, a {\it subspace} of gauge invariant states
those correspond to the compact closed surfaces are allowed to gauge
transform into each other. Thus, further restrictions are imposed on
this class of quantum states, \cite{Ashtekar-surface}. The quantum
states of a compact closed underlying geometry yields to those that
satisfy the two conditions on the side bulk edge spins:
$\sum_{\alpha} j_u^{(\alpha)} \in \mathbb{Z}^+$ and $\sum_{\alpha}
j_d^{(\alpha)} \in \mathbb{Z}^+$. However, due to the existence of
sum in these conditions, the spin of the majority of bulk edges in
the near-horizon region are left unconditional. In other words, the
conditioned trivalent states among all ingredient states of whole
surface state is one or a few.

The quantum surface that is associated to a black hole horizon
semi-classically determines the quantum decays of energy from the
black hole. This definition is only restricted to the case of black
holes and does not hold in any random surface. In the rest of the
note the spectroscopy of the decays is illustrated. Before it, in
the next part, an important symmetry within the eigenvalues are area
operator is uncovered.


\part{Part I: The symmetry of area spectrum} \label{part1}

\markboth{\sc \ssz Part I: The symmetry of area spectrum}{\sc \ssz
M. Ansari}

In this part, by the use of number theory a significant property of
the area eigenvalues is uncovered. Having known the complete
spectrum of area symmetry (\ref{eq. A degenerate sec.}) a reduced
formula is written. As a consequence, the complete spectrum of area
eigenvalues in both group $SU(2)$ and $SO(3)$ representations can be
split into the mixtures of equidistant numbers. This lead to the
quantum amplification effect, which is described in next part.


\section{SO(3) area and Square-free
numbers}\label{subsec_SO(3)_and_square_free}

In $SO(3)$ group representation, the spins are positive integers.
Evaluating $\frac{1}{2}\left(m_{j_u, j_d,j_{u+d}}\right)^2$, if all
repetitions of numbers (degeneracies) are identified, the whole {\it
Natural} numbers are reproduced. This is proved in the Appendix
(\ref{app1}).

As an immediate consequence, there exists a irreducible formula for
the eigenvalues of area which depends only on one integer number.
The irreducible formula of the complete area eigenvalues is

\begin{equation}\label{eq. irreduc area}
    a_n = 4 \pi \gamma \ell_P^2 \chi\  \sqrt{n},
\end{equation}
where $\chi = \sqrt{2}$ and $n \in \mathbb{N}$.

In fact the eigenvalues of area operator in the original formula
(\ref{eq. A degenerate sec.}) that depends on three variables $j_u$,
$j_d$, $j_{u+d}$ is a reducible representation of the set. If
degeneracies are identified the irreducible formula (\ref{eq.
irreduc area}) appears.

Any integer is the multiplication of a `square-free' number and a
square number.\footnote{The proof is in Appendix (\ref{app1}).} By
definition, an integer is said to be square-free, if its prime
decomposition contains no repeated factors. For example, $30$ is
square free since its prime decomposition $2\times3\times5$ contains
no repeated factors. Consider the natural number $2^{5} \times
3^{8}$. This number can be rewritten in the form $(2)\times (2^2
\times 3^4)^2$. The first part is a square-free number and the
second one is a square number.

Having this decomposition of natural numbers, consider the sequence
of numbers containing the same square-free factor multiplied by all
squared numbers. For example the sequence $\{$ 3, 12, 27, 48, 75,
108, 147, $\cdots \}$, which is in fact $3\times n^2$ for $n \in
\mathbb{N}$. This sequence can be written in the form $3\times \{$1,
4, 9, 16, 25, 36, 49, $\cdots\}$, or briefly $3 \mathbb{N}^2$. Such
a sequence of numbers is called a squared set. The square-free
number $3$ based on which the sequence $3\mathbb{N}^2$ is produced
is the
 {\it representative} of the squared set $3\mathbb{N}^2$. We indicate the representative with the
symbol $\zeta$ and its corresponding square generation with $\zeta
\mathbb{N}^2$.

Obviously, taking square root from the elements of a squared
generation, say $\zeta \mathbb{N}^2$, an equidistant sequence of
numbers is produced, $\sqrt{\zeta} \mathbb{N}$. This evenly spaced
set of numbers is called a `{\it generation}.' In fact, by doing
this we decompose the set of numbers $\sqrt{n}$ into $\sqrt{\zeta}
m$ for integer $n$ and $m$ and square-free $\zeta$. Consequently,
the formula (\ref{eq. A degenerate sec.}) is performed into the
following reducible but important form:
\begin{equation}
 \label{eq. a eta m = ...} a_n\left(\zeta\right) =  (4
\pi \gamma \ell^2_{\mathrm{P}} \chi)\ n \sqrt{\zeta} ,
\end{equation}
where $\chi = \sqrt{2}$ $n \in \mathbb{N}$, and $\zeta \in
\mathbb{A}$ for $\mathbb{A}$ stands the set of square-free numbers.
The list some of the square-free numbers are given in Table
(\ref{tab. square-free}) of Appendix (\ref{app1}).

What is special about this final formula is that it represents
clearly that a generation with representative $\zeta$ gets evenly
spaced area eigenvalues.

A curious reader is encouraged to read more details in the Appendix
(\ref{app1}).


\section{SU(2) area and positive definite quadratic forms}
\label{subsec_SU(2)_and_discriminant}

In $SU(2)$ group representation, evaluating $4
\left(m_{j_u,j_d,j_{u+d}}\right)^2$ from (\ref{eq. A degenerate
sec.}) produces the congruent numbers unto 0 or 3 mod 4. The proof
is in the Appendix (\ref{app2}). These numbers are the page numbers
of a book that is printed out only at the pages that come after each
two leaves of sheets by face, 3, 4, 7, 8, 11, 12, 15, 16, 19, 20,
etc. These numbers are also called {\it the Skew Amenable numbers},
\cite{Barger}.

Since the Skew Amenable number cannot be fitted into a formula with
one variable. Instead, it can be fitted into the combination of
these two sets: $\left( 4 \pi \gamma \ell^2_{\mathrm{P}} \chi
\right)\ \sqrt{ 4n}$ and $\left( 4 \pi \gamma \ell^2_{\mathrm{P}}
\chi \right)\ \sqrt{ 4n+3}$, where $\chi=1/2$ and $n \in
\mathbb{N}$. It can be proven that any skew amenable number $b'$ can
be written in terms of $b \times n^2$ for $n \in \mathbb{N}$. The
numbers $b$ are the elements of a subset of Skew Amenable numbers,
the subset $\mathbb{B}$, that contains the discriminants of every
positive definite quadratic forms.\footnote{For proofs refer to the
Appendix (\ref{app2}).}

Henceforth, the complete spectrum of area eigenvalues
$m_{j_u,j_d,j_{u+d}}$ is equivalent to the family of the generations
$\{(\sqrt{\zeta}/2) \mathbb{N}\}$, where $\zeta \in B$. Area
eigenvalues, instead of being determined by three quantum numbers
$j_u$, $j_d$, and $j_{u+d}$, can be performed by two as
 \begin{equation}
 \label{eq. a eta m = ...2}
 a_n\left(\zeta\right) = \left( 4
\pi \gamma \ell^2_{\mathrm{P}} \chi \right)\ n \sqrt{ \zeta},
\end{equation}
where $n \in \mathbb{N}$ and $\chi = 1/2$. The list of some of the
discriminants is given in Table {\ref{tab. B}) of Appendix
(\ref{app2}).

A curious reader is encouraged to read more details in the Appendix
(\ref{app2}).

\section{Conclusion}
\label{subsec_SU(2)_and_discriminant}

    Remarkably the area eigenvalues in a reduced form in both group representations $SU(2)$ and $SO(3)$ are performed into one formula.
    In the above two subsections it was justified that the
complete spectrum of area operator indeed can be specified by two
indices $n$ and $\zeta$, instead of three indices $j_u, j_d$, and
$j_{u+d}$,
\begin{equation}
 \label{eq. a both} a_n{\left(\zeta\right)} = \left( 4  \pi \gamma \ell^2_{\mathrm{P}}
\chi \ \right)\ n \sqrt{\zeta},
\end{equation}
where $n \in \mathbb{N}$.  In $SO(3)$ representation, $\zeta \in
\mathbb{A}$ and $\chi = \sqrt{2}$. In $SU(2)$ representation, $\zeta
\in \mathbb{B}$ and $\chi = 1/2$. $\chi$ is the
 {\it group characteristic parameter} and $\zeta$ is the  {\it generation representative}. Therefore, in both group
representations, the area eigenvalues exhibit equally spaced
symmetry which make one of the original labels redundant.

Having defined the area eigenvalues, the following Lemmas can be
easily investigated:

 {$\blacktriangleright$} {\bf Lemma
1}: {\it Having two eigenvalues $a_1\in \sqrt{\zeta_1} \mathbb{N}$,
and $a_2 \in \sqrt{\zeta_2}\mathbb{N}$, where $\zeta_1 \neq
\zeta_2$, for any choice of the eigenvalues in the corresponding
generations these two eigenvalues are not equal, $a_1 \neq a_2$.}
 {$\blacktriangleleft$}

 {$\blacktriangleright$} {\bf Lemma
2}: {\it Having two eigenvalues $a_1\in \sqrt{\zeta_1} \mathbb{N}$,
and $a_2 \in \sqrt{\zeta_2}\mathbb{N}$, where $\zeta_1, \zeta_2 \in
\mathbb{A}$ (or $\mathbb{B}$) and $\zeta_1 \neq \zeta_2$, there in
no eigenvalue in any generation that is equal to $a_1\pm a_2$.}
 {$\blacktriangleleft$}


\part{Part II: Radiation} \label{sec_radiation}
\markboth{\sc \ssz Part II: Radiation}{\sc \ssz M. Ansari}

In this part, based on the results of the Part (I), the spectroscopy
of a quantum black hole is worked out.

Let us briefly overview the rest of this Part. The quantum
fluctuation of the horizon area of a Schwarzschild black hole may
occurs at the state associated to one or more than one of the
horizon area cell. Since in the complete spectrum of area the gap
between consecutive eigenvalues decreases in larger eigenvalues,
effectively a continuous set of radiance frequencies are expected.
Considering the result of Part (I), in which the complete spectrum
of area is uncovered to exhibits evenly spacing symmetry if it is
classified into some subsets (the so-called `generations'). Consider
a transition from an upper area level, which belongs to the
generation $\zeta_1$, into a lower area level, which belongs to the
generation $\zeta_2$. While there is nothing special with transition
between two levels of two different generations, the radiance
intensities of a set of frequencies which correspond to the
transition within a generation ($\zeta_1 = \zeta_2=:\zeta$) get
highly amplified. The reason is that within a generation of quantum
area a typical transition can occur from many different levels. For
instance, a quantum leap of the scale of the double of the gap
between a generation can be initiated from the third, fourth, fifth,
up to the maximum levels. These quanta are all different copies of
the same energy that a black hole may radiate. In fact, quantum
amplification results into discrimination between the spectral line
intensities. Such emissions are unblended and amenable to possible
observation in primordial black holes.

Considering the symmetry of area each one of the generations
justifies the equidistant ansatz (\ref{eq. Bek ansatz}) separately
after replacing $\alpha = 4  \pi \gamma \chi \sqrt{\zeta}$. The
fundamental frequencies which are emitted by quantum leap inside a
generation $\zeta$ is
\begin{equation}\label{eq. w^eta = }
  \varpi\left(\zeta\right) = \frac{ \gamma c^3}{8 G M} \chi
\sqrt{\zeta}.
\end{equation}

Let us name $\omega_o := c^3/8GM$
  {\it the frequency scale factor} and is of
 the order of $10^{16}/M_{kg}$ (eV).  For instance the frequency scale corresponding to a black hole of mass $M \sim
10^{12}$ kg is of the order of 10 keV and thus the harmonics are of
order $10\sqrt{\zeta} n$ keV, though these lines are not of the same
intensities. This frequency if is associated to a primordial black
hole it is subject to redshifting of the order of three order of
magnitude. In fact, the intensity is suppressed as the gap between
the levels of a generation grows. This because of the amount of
quantum amplification a frequency may take. A precise work based on
minimal number of natural assumptions is required to work out the
intensities, which is introduced in the rest.


\section{Quantum amplification effect}\label{sec decay}
In general, transitions fall into two categories: ($i$)
 {\it the generational transitions},
quantum leaps from a level to a lower level of the same generation,
and ($ii$)  {\it the inter-generational transitions}, quantum leaps
from a level into a lower level of different generation. The
frequency corresponding to the first type of emissions is
proportional to the fundamental frequency of the generation to which
both the initial and final levels belong, $\omega_n := n
\varpi(\zeta)$ for integer positive $n$. These frequencies are
called  {\it harmonic frequencies} of the generation $\zeta$.
Whatever frequency which is not of this type is of
 {\it non-harmonics}.

The strategy of determining the intensity of radiation is as
follows. The intensity of an emissive frequency is defined by the
amount of energy radiating at that frequency per unit time and area.
The energy corresponding to a frequency is proportional to the
average number of its emissive quanta. Firstly, it is assumed that
the emissions occur in sequences. Accordingly, the probability of
emissions of a typical sequence is determined. Having this, one can
calculate the probability of the sequence that contains a number of
the same frequencies. The average number of emissive quanta at
different frequencies are determined. Thus, the intensity of
frequencies are found. We calculate the intensity and the natural
width of lines and the corresponding temperature to a black hole in
this section. For the matter of clarifying the hidden assumption
behind this strategy we give main axioms individually.
\\

\mbox{ {\it Axiom 1: Emissions occur in a sequential order.}}
\\

This was first proposed by Ulrich Gerlach at the surface of a
collapsing star in \cite{Gerlach}.  From a black hole, as a possible
ultimate state of a collapsing star, a quantum of energy may be
emitted between two classical stationary states. Describing the
decay of the black hole during any interval of observer time $\Delta
t$, a set of $j$ individual decays are emitted in the sequence
$\{\omega_1, \omega_2, \cdots, \omega_n\}$, successively. The
probability of a typical sequence is determined in this section.

In generational transitions, many copies of a harmonic frequency can
be produced from different pairs. However, this is not the case for
non-harmonics, because the irrational numbers $\sqrt{\zeta}$ that
area eigenvalues are proportional to, cannot be decomposed into a
sum of other irrational numbers. Accordingly, Lemma 2 approves that
the difference $\Delta a = a(\zeta)-a'(\zeta')$ between two levels
of different generations, $\zeta \neq \zeta'$, is `unique' and
cannot be produced by considering other pairs. Therefore, a
non-harmonic transition is emitted only from
 {\it one} pair of levels.

On the other hand, from the classical relation between the horizon
and the black hole mass (\ref{eq. A=M^2}), it is easy to verify
$A(\mathrm{m}^2) = 2.77 \times 10^{-53} M({\mathrm{kg}})^2$. The
temperature of such a black hole is $T(\mathrm{k})=1.23 \times
10^{23} /M(\mathrm{kg})$. For instance, the horizon area
corresponding to the black hole of mass $10^{12}$ kg is $2.77 \times
10^{-29} \mathrm{(m^2)}$ and its temperature is $1.228 \times
10^{11}$k. Such a horizon is 40 order of magnitude larger than the
quanta of area! This gives the confidence that the number of levels
that contribute to the radiation procedure is enormous.

This fact make the difference between harmonics and non-harmonics
important. Namely, the population of harmonics exceed the population
of non-harmonics. This effect in quantum mechanics is called
 {\it Quantum Amplification Effect}.
This effect has a strong root in the symmetry of area.

To determine how much the difference of the population is important
and if it is visible a precise analysis is necessary. Let us start
off the analysis with the probability of some decay in a sequential
order.


\subsection{The probability of time-ordered decays}

The probability of one jump (no matter of what frequency) in the
course of time $\Delta t$ is shown by $P_{\Delta t}(1)$. Similarly,
the probability of no jump is $P_{\Delta t}(0)$. During the time
interval $2 \Delta t$, the probability of no jump (the failure of
decaying) is equal to the probability of the failure in each one of
its two fragment of time intervals, $P_{2 \Delta t} (0) = [P_{\Delta
t} (0)]^2$.  The general solution of this functional equation is
$P_{\Delta t} (0) = \exp (- \Delta t / \tau)$, where $\tau$ is the
survival timescale of the black hole from decaying.

We let the horizon to decay a sequence of frequencies successively.
Using the same argument, the probability of one jump (of any
frequency) in time interval $2 \Delta t$ is $P_{2 \Delta t} (1) = 2
P_{ \Delta t} (0)P_{\Delta t} (1)$. Therefore, $P_{ \Delta t} (1) =
(\Delta t / \tau') \exp(- \Delta t / \tau) $.

The probability of 2 jumps in the time interval $2\Delta t$ can be
written as $P_{ 2\Delta t} (2)=2P_{ \Delta t} (0)P_{ \Delta t} (2) +
[P_{ \Delta t} (1)]^2$. This formula can be extended to the
probability of emission of `even' number ($j$) quanta in the time
interval $2\Delta t$,

 \beq \label{eq. even j} P_{ 2\Delta t} (j)=2
\sum_{i=0}^{i=\frac{j}{2}-1}P_{ \Delta t} (i)P_{ \Delta t} (j-i) +
[P_{ \Delta t} (j/2)]^2. \eeq

On the other hand, the probability of 3 jumps in the time interval
$2\Delta t$ can be written as $P_{ 2\Delta t} (3)=2P_{ \Delta t}
(0)P_{ \Delta t} (3) + 2P_{ \Delta t} (1)P_{ \Delta t} (2)$. This
formula can deduce to the probability of emission of `odd' number
($j$) quanta in the time interval $2\Delta t$,

\beq \label{eq. odd j} P_{ 2\Delta t} (j)=2
\sum_{i=0}^{i=\frac{j-1}{2}}P_{ \Delta t} (i)P_{ \Delta t} (j-i).
\eeq

Generating all the probabilities starting from $P_{\Delta t}(0)$ up
to $P_{\Delta t}(j-1)$ consecutively from the recursive formula
(\ref{eq. even j}) and (\ref{eq. odd j}), one can generate the
probability of $j$ emissions as a function of $j$ and $\Delta t$.
The general solution for the probability of $j$ decays is
\begin{equation}\label{eq. P(j)=}
P_{ \Delta t} (j) = \frac{1}{j!} \left( \frac{\Delta
t}{\tau'}\right)^j \exp(- \Delta t / \tau ),
\end{equation}
in which by normalization $\tau' = \tau$.


\subsection{The probability of a decay}

Consider a sequence of radiance frequencies $\{\omega_1, \omega_2
\cdots \}$. These frequencies might be harmonics or non-harmonics.
For the purpose of determining the probability of this sequence let
us begin with one jump ($j=1$) in the course of time $\Delta t$.

Since the emissions are supposed to occur in time order, the
probability of a sequence of decays is the product of conditional
probability and the probability distribution of time ordering. Thus,
the probability of one emission is $P_{\Delta t}(\{\omega\}) =
P_{\Delta t}(\{\omega\ | \ 1\})P_{\Delta t}(j=1)$. This probability
is not difficult to be determined. Before this the second axiom is
introduced.
\\

\mbox{  {\it Axiom 2: The entropy of a Schwarzschild black hole is
dominantly  $A/4\ell_P^2$.}}
\\

Entropy is defined as the logarithm of the number of microstates of
a macroscopic state. Since the macroscopic state of a Schwarzschild
black hole is determined by one parameter, the horizon area, the
entropy associated to a black hole of horizon area $A$ is determined
by the number of quantum states associated to such a horizon. This
degeneracy, $g(A)$, is dominantly $g(A) = \exp(A/
4\ell^2_{\mathrm{P}})$. The horizon area of a loop quantum black
hole is made of $N$ patches of area eigenvalues, $A = \sum_{i=1}^N
a_i$. Therefore the black hole degeneracy is in fact
$g(A)=\exp(\sum_{i=1}^N a_i/ 4\ell^2_{\mathrm{P}})$. On the other
hand, the overall degeneracy $g$ corresponding to a system that is
made of $N$ subsystems each with individual degeneracy $g_i$, is
$g=\prod_{i=1}^{N}g_i$. Due to the Lemma 2, the contribution of each
generation in the horizon area $A$ cannot be replaced by the
combination of area levels of the other generations. Therefore, the
macroscopic horizon area is split into its ingredient area
contribution of each generation, $g(A)=\prod_{\zeta}\exp(\sum_{i}
a_i(\zeta)/ 4\ell^2_{\mathrm{P}})$. Therefore, the degeneracy
associated to the generation $\tau$ is $g(\zeta)=\exp(\sum_{i}
a_i(\zeta)/ 4\ell^2_{\mathrm{P}})$, where $i$ indicates the levels
of the generation that contribute in the horizon area. Since each
generation is equidistant, all level that contribute in the horizon
area from one generation sum into a level inside the same
generation, say the level $n$. In other words, $\sum_{i}a_i(\zeta)
=: a_n(\zeta)$. Consequently, the degeneracy $g(\zeta)$ can be
thought of being the degeneracy associated to the level $n$ of the
generation, $g(n; \zeta)=\exp(a_n(\zeta)/ 4\ell^2_{\mathrm{P}})$. By
the use of equation (\ref{eq. a both}), the degeneracy of a typical
level $a_n(\zeta)$ is
\begin{equation}\label{eq g()}
g(n; \zeta) := e^{ \pi \gamma \chi \sqrt{\zeta} \ n}.
\end{equation}
where $\zeta$ is the representative number of the generation, $n$ is
the level of the frequency in the generation ladder. \footnote{This
degeneracy can also be determined exactly by considering the fact
that a typical area level $a_n(\zeta)$ can be made of some smaller
area patches (of the same generation) in some different
configurations. Considering the degeneracy of each level, the
exponentially growing degeneracy of a level is immediately
reproduced, \cite{Ansari2006locality}.}

An area patch of level $n$ of the generation $\zeta$ may decay into
the level $n'$-th of the generation $\zeta'$, where $\sqrt{\zeta} n
> \sqrt{\zeta'} n'$. In this process, the degeneracy $g(n;
\zeta)$ changes into $g(n'; \zeta')$. By the use of (\ref{eq g()}),
the transition into a lower level changes the degeneracy by a factor
of $\exp( - \pi \gamma \chi | \sqrt{\zeta} n - \sqrt{\zeta'} n'|)$
corresponding to the emission of the frequency $\omega = (\gamma
\chi c^3/ 8 GM) (\sqrt{\zeta} n - \sqrt{\zeta'} n')$.

The `population' of a quantum of area is the number of different
pairs of levels that produce it. This number can be normalized to
one by the use of the maximum population number, $N_o$, which is in
fact nothing but the number of level pairs that produce the
fundamental frequency of the first generation (the generation whose
corresponding gap between levels is minimal). Therefore, the
 {\it population weight} of the
frequency $\omega$ is defined as $\rho(\omega) = N/N_o$, where $N$
is the number of pairs that produce the frequency $\omega$. It is
also clear that the population weight of non-harmonics is $1/N_o$.
\begin{figure}[h]
\begin{center}
\includegraphics[width=10cm]{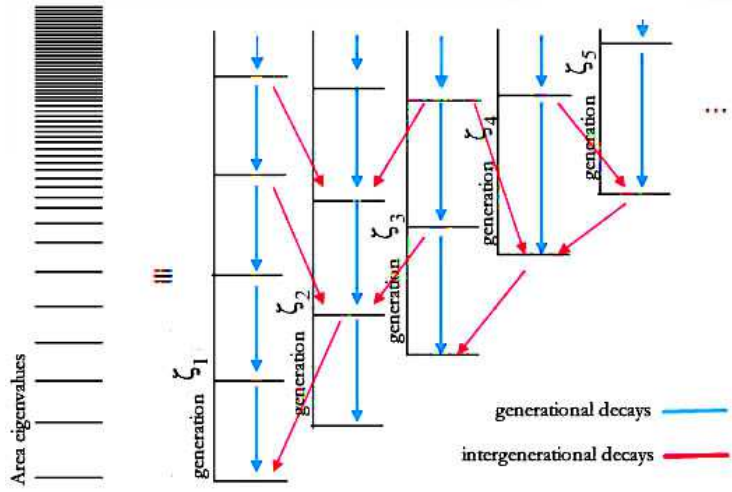}
\end{center}
\caption{A schematic diagram of generational emissions (the vertical
black arrows) and inter-generational emissions (the slanted red
arrows) for a few generations.} \label{cap. emissions}
\end{figure}

In our navigation for determining the probability of a specific
frequency emission the third axiom is introduced:
\\
\begin{quote}

\mbox{
  {\it Axiom 3: The density matrix elements for quantum transitions}}

  \mbox{{\it \ \ \ \ \ \ \ \ \ \ \ \ \ \ \ \ \ \ between near levels are uniform.}}
\\
\end{quote}

Having assumed this, the probability of a jump is proportional to
the change of degeneracy as well as the population weight of the
frequency. Therefore, the conditional probability of a typical
frequency $\omega$ by the use of (\ref{eq g()}) is

\begin{equation}\label{eq. prob(general freq | 1)}
   P_{\Delta t}(\{\omega \}\ | \ 1) = \frac{1}{C} \rho({\omega}) \ e^{ - \Lambda
   \omega},
\end{equation}
where \beq \label{eq. gamma} \Lambda:=8 \pi  G M / c^3. \eeq

The normalization relation of the probability determines $C$. It is
defined to be $C := \sum_{\omega} \rho({\omega})\ e^{ - \Lambda
\omega}$.

Since the probability of a typical frequency $\omega$ is determined
from $P_{\Delta
 t}(\{\omega\}) = P_{\Delta
t}(\{\omega\}\ |\ 1)\ P_{\Delta t}(1)$. By the use of (\ref{eq.
P(j)=}) and (\ref{eq. prob(general freq | 1)}) for $j=1$,
\begin{equation}\label{eq. Pw}
P_{\Delta t}(\{\omega\}) = \frac{\Delta t }{C \tau}
 e^{-
\Delta t / \tau}\ \rho(\omega)\ e^{-\Lambda \omega}
\end{equation}

In the case of generational decays, the decay condition
$\sqrt{\zeta} n
> \sqrt{\zeta'}  n'$ is reduced into $n > n'$. The conditional
probability of a typical frequency $\omega_m := m \varpi(\zeta)$
emission, where $m = n-n'$, by the use of (\ref{eq. w^eta = }) and
(\ref{eq. prob(general freq | 1)}) reads
\begin{equation}\label{eq. pw|1}
P_{\Delta t}(\{\omega_m(\zeta)\}\ | \ 1) = \frac{1}{C} \
\rho(\zeta)\ q(\zeta)^{-m},
\end{equation}
where $q(\zeta) := e^{\Lambda \varpi(\zeta)}$ is independent of mass
and dependent to the generational representative number $\zeta$. In
fact it is \footnote{Comparing (\ref{eq. definition of q}) with the
degeneracy of a level (\ref{eq g()}) shows the simple relation:
$g(n; \zeta) = q(\zeta)^n$.}

\beq \label{eq. definition of q} q(\zeta) := e^{\pi \gamma \chi
\sqrt{\zeta}}. \eeq

Given a black hole of horizon area $A$, it is discussed in section
(\ref{sec decay}) a generation with smaller gap between levels
produces more copies of each one of its corresponding harmonic
frequencies. Since the gap between the levels is $(4 \pi \gamma \chi
\sqrt{\zeta})\ \ell^2_{\mathrm{P}}$ by the use of (\ref{eq. a both})
the number of levels below the horizon area is $N:=A/(4 \pi \gamma
\chi \sqrt{\zeta}) \ell^2_{\mathrm{P}}$, which is a huge number
(about $10^{40}$ levels). On the other hand, the number of $m$-level
jumps down a ladder of total $N$ levels is $N-m$. In a classical
black hole $N$ is extremely large, henceforth the population weight
of the harmonics of frequency $\omega_m(\zeta)$ for $m \ll N$ is
$\rho(\zeta) = N/N_o = \sqrt{\zeta_o/\zeta}$, where $\zeta_o$ is the
generation with the minimal gap between levels.\footnote{$\zeta_o$
in $SU(2)$ version is 3 and in $SO(3)$ version is 1.} Dropping the
constant coefficient $\sqrt{\zeta_o}$ from the definition, the
population weight of small harmonics is
\begin{equation} \label{eq. rho}
\rho(\zeta) := \frac{1}{\sqrt{\zeta}}.
\end{equation}

Since the probability of the harmonic frequency $\omega_n(\zeta)$ is
defined by $P_{\Delta
 t}(\{\omega_n(\zeta)\}) = P_{\Delta
t}(\{\omega_n(\zeta)\}\ |\ 1)\ P_{\Delta t}(1)$. By the use of
(\ref{eq. pw|1}) and (\ref{eq. P(j)=}) for $j=1$, one can write the
probability of the harmonic frequency
\begin{equation}\label{eq. Pw}
P_{\Delta t}(\{\omega_n(\zeta)\}) = \frac{\Delta t }{C \tau}
 e^{-
\Delta t / \tau}\ \rho(\zeta)\ q(\zeta)^{-n}
\end{equation}

What is $C$? In fact after a moment of analytical calculation the
normalization coefficient $C$ is found

\begin{equation}\label{eq. C}
    C = \sum_{\mathrm{all}\ \zeta} \ \frac{\rho(\zeta)}{q(\zeta)-1}
\end{equation}

It is easy to prove that $C$ is a finite number of the order $O(1)$.
A curious reader is encouraged to read the detail of the derivation
of $C$ and testing its finiteness in Appendix (\ref{app4}).

Next step is to generalized this probability for a sequence of $j$
successive emissions of different frequencies.


\subsection{The probability of a sequence of emissions}}
Following the Axiom 1, the generalized probability of a sequence of
harmonics is $P_{\Delta t}(\{ \omega_1, \omega_2, \cdots, \omega_j\}
)$, where the frequencies can be harmonics or non-harmonics. Let us
assume the time interval is made of $S$ fragments of smaller time
intervals, $\Delta t = S \epsilon $, where $S \gg j$ and each one of
the $j$ decays occurs in one fragment of time $\epsilon$. There are
$S!/j!(S-j)!$ number of ways for selecting $j$ jumping intervals out
of total $S$ time intervals. This number of ways for the case of $S
\gg j$ is approximated to $S^j/j!$. In the overall $j\epsilon$
moment intervals out of $S$ ones, the black hole successfully decays
and in the rest of time, $(S - j )\epsilon$, it fails to decay. The
probability of $j$ emissions is thus $ (S^j/j!)
P_{\epsilon}(0)^{S-j}\ \prod_{i=1}^j P_{\epsilon}(\{ \omega_i\}) $.
Substituting $P_{\epsilon}(\{\omega_i\})$'s from (\ref{eq. Pw}), the
probability is:
\begin{equation}\label{eq. P(n1, n2 ...non)=}
P_{\Delta t} (\{\omega_1, \omega_2, \cdots, \omega_j \}) =
\frac{1}{j!} \left(\frac{\Delta t}{C\tau}\right)^j \ e^{-\Delta t /
\tau} \ \prod_{i=1}^{j} \rho(\zeta_i)\ e^{-\Lambda \omega(\zeta_i)}.
\end{equation}

In the presence of $r$ non-harmonics in a sequence of frequencies
decreases the probability of the sequence by a factor of
$(1/N_o)^r$, which is negligible for classic black holes. In fact,
only the harmonics take a major contribution to determining the
intensities.

Let us consider now a sequence of harmonic emissions of different
generations, $\{\omega_{n_1}(\zeta_1), \omega_{n_2}(\zeta_2),
\cdots, \omega_{n_j}(\zeta_j)\}$. According to (\ref{eq. pw|1}) and
(\ref{eq. P(n1, n2 ...non)=}) the probability of the sequence is
\begin{equation}\label{eq. P(n1, n2 ...)=}
P_{\Delta t} (\{\omega_{n_1}(\zeta_1), \omega_{n_2}(\zeta_2),
\cdots, \omega_{n_j}(\zeta_j)\}) = \frac{1}{j!} \left(\frac{\Delta
t}{C\tau}\right)^j \ e^{-\Delta t / \tau} \ \prod_{i=1}^{j}
\rho(\zeta_i)\ q(\zeta_i)^{-n_i}.
\end{equation}

By the use of the probability of $j$ decays in the course of time
$\Delta t$ from (\ref{eq. P(j)=}), the generalized conditional
probability of a sequence of harmonics is found
\begin{equation}\label{eq. P(n1, n2 ...|j)=}
P_{\Delta t} (\{\omega_{n_1}(\zeta_1), \omega_{n_2}(\zeta_2),
\cdots, \omega_{n_j}(\zeta_j)\} \ | \ j) =
\left(\frac{1}{C}\right)^j\ \ \prod_{i=1}^{j} \rho(\zeta_i)\
q(\zeta_i)^{-n_i}.
\end{equation}

This conditional probability turn out to be independent of time.

Since the intensity of a harmonic frequency depends on the average
number of the emission in the course of time. This average number
depends on the probability of $k$ emissions of the emissive
frequency in any sequence of dimension $j\geq k$.


\subsection{The probability of $k$ quanta of the same frequency}

Let us assume that among the $j$ emissions there exist $k$ quanta of
the frequency $\omega_{n_k}(\zeta_k)$ and the rest $j-k$ frequencies
belong to other frequencies. Consider the $j$ dimensional sequence
$\{\omega_{n_1}(\zeta_1), \omega_{n_2}(\zeta_2), \cdots,
\omega_{n_k}(\zeta_k), \cdots , \omega_{n_k}(\zeta_k), \cdots,
\omega_{n_j}(\zeta_j) \}$ in which there are $k$ quanta of the same
frequency $\omega_{n_k}(\zeta_k)$. If the black hole makes $j$
decays such that $k$ of them are of the same frequency
$\omega_{n_k}(\zeta_k)$, (for $k \leq j$), there are $k!/j!(j-k)!$
ways to select these $k$ quanta. The probability of each selection
due to (\ref{eq. P(n1, n2 ...|j)=}) is

\beq \label{eq. p(...k..k..)} \left(\frac{1}{C} \right)^{j}\
\left(\rho\left(\zeta_k\right) q\left(\zeta_k\right)^{-
n_k}\right)^k \prod_{i=1}^{j-k} \rho(\zeta_i)\ q(\zeta_i)^{-n_i}.
\eeq where in the product part of it the frequencies are any
frequency except $\omega_{n_k}(\zeta_k)$.

For the purpose of determining the probability of $k$ emissive
quanta of the frequency $\omega_{n_k}(\zeta_k)$ in a $j$ dimensional
string included all possible accompanying frequencies, $P_{\Delta t}
(k \ |\ \omega_{n_k}(\zeta_k) , j)$, we should sum over the
probabilities (\ref{eq. p(...k..k..)}) for all possible frequencies
associated to the accompanying frequencies, all frequency values
except $\omega_{n_k}(\zeta_k)$. Since the non-harmonic emissions are
continuous sum over non-harmonics is effectively evaluated by
integral. We must consider different cases for the $j-k$
accompanying emissions: the case that none of the $j-k$ frequencies
is non-harmonic, the case that only one of them is non-harmonic,
etc..  Therefore the conditional probability is
\begin{eqnarray*}\label{eq. P(k|nw)}
  P_{\Delta t} (k  |
\omega_{n_k}(\zeta_k) , j)
 = &&\frac{j!}{ k! (j-k)!} \left(\frac{1}{C}  \right)^{j}  \left(\rho\left(\zeta_k\right) q\left(\zeta_k\right)^{-
 n_k}\right)^k\times  \\
& & \left[ \prod_{i=1}^{j-k}\ \sum_{\mathrm{all}\ \zeta}
\sum_{\omega \neq \omega_{n_k}} \rho(\zeta_i)\ q(\zeta_i)^{-n_i}
\right. \\ && + \prod_{i=1}^{j-k-1}\ \sum_{\mathrm{all}\
\zeta}\sum_{\omega \neq \omega_{n_k}} \rho(\zeta_i)\
q(\zeta_i)^{-n_i}\left(\int_{0, \omega \neq \omega_k}^{\infty} \rho
e^{-\Lambda \omega}
\Lambda d\omega \right)\\
&& + \prod_{i=1}^{j-k-2}\ \sum_{\mathrm{all}\ \zeta}\sum_{\omega
\neq \omega_{n_k}} \rho(\zeta_i)\ q(\zeta_i)^{-n_i}\left(\int_{0,
\omega \neq \omega_k}^{\infty} \rho e^{-\Lambda \omega}
\Lambda d\omega \right)^{2} +\cdots\\
&& + \sum_{\mathrm{all}\ \zeta}\sum_{\omega \neq \omega_{n_k}}
\rho(\zeta_i)\ q(\zeta_i)^{-n_i}\left(\int_{0, \omega \neq
\omega_k}^{\infty} \rho e^{-\Lambda \omega}
\Lambda d\omega \right)^{j-k-1}\\
&& + \left.\left(\int_{0, \omega \neq \omega_k}^{\infty} \rho
e^{-\Lambda \omega}
\Lambda d\omega \right)^{j-k}\right]. \\
\end{eqnarray*}

Substituting $\rho$, the contribution of a non-harmonic emission
becomes $(1/N_o)(1-e^{-\Lambda \omega_k})$. By the use of the
equality (\ref{eq. C}) the sum $\sum_{\mathrm{all}\ \zeta}\sum_{n
\neq n_k} \rho(\zeta) q(\zeta)^{-n}$ gives rise to  $C -
\rho(\zeta_k)q(\zeta_k)^{-n_k}$.


In the classical limit, $ 1/N_o \rightarrow 0$, inside the bracket
all terms with the factor $1/N_o$ are higher order corrections to
the probability. For the purpose of determining the intensity, it is
sufficient to consider only the first order term. Effectively this
probability is

\begin{eqnarray}\label{eq. p(w|k,j)}
  P_{\Delta t} (k \ |\
\omega_{n_k}(\zeta_k), j)
 = \frac{j!}{ k! (j-k)!} \left(\frac{1}{C} \right)^{j}\ \left(\rho\left(\zeta_k\right) q\left(\zeta_k\right)^{-
 n_k}\right)^k\ \nonumber \\
\times \left( C -
\rho\left(\zeta_k\right)q\left(\zeta_k\right)^{-n_k}\right)^{j-k}.
\end{eqnarray}

We multiply this conditional probability by the absolute probability
distribution $P_{\Delta t} (j)$ in equation (\ref{eq. P(j)=}) and
sum over all $j\geq k$ in order to provide the probability of
$P_{\Delta t}\left( k | n_k \varpi^{\left(\zeta\right)} \right)$,

\begin{eqnarray*}\label{eq. P(k|nw)}
  P_{\Delta t} (k \ |\
\omega_{n_k}(\zeta_k))
 =
\frac{1}{ k!} && \left(\rho\left(\zeta_k\right)
q\left(\zeta_k\right)^{- n_k}\right)^k \ \left( C -
\rho\left(\zeta_k\right)q\left(\zeta_k\right)^{-n_k}\right)^{-k} e^{- \Delta t / \tau } \\
&& \times \sum_{j \geq k}^{\infty}  \frac{1}{(j-k)!}
\left(\frac{\Delta t}{C \tau}\right)^j \left( C -
\rho\left(\zeta_k\right)q\left(\zeta_k\right)^{-n_k}\right)^{j}
\end{eqnarray*}

Applying the equality $\sum_{a=b}^{\infty} z^a/(a-b)! = z^b \exp(z)$
by replacing $a,b$ with $j, k$ respectively in the second line, the
probability distribution of the emission $k$ quanta of frequency
$\omega_{n_k}(\zeta_k)$ is determined,

\begin{equation}\label{P(k|nw)}
  P_{\Delta t}\left( k\ |\ \omega_{n_k}(\zeta_k) \right) = \frac{1}{k!}
  \left( x_{n_k}\left(\zeta_k\right)\right)^{k} e^{-x_{n_k}\left(\zeta\right)},
\end{equation}
where $x_n\left(\zeta\right):= (\Delta t / C\tau)\ \rho(\zeta)
q\left(\zeta\right)^{-n}$. This probability turns out to be
Poisson-like
distribution.


\section{Intensity} By definition, the intensity of
$\omega_{n_k}(\zeta_k)$ is the total energy that is emitted at this
frequency and unit time per unit area. Since the emissions of
diverse frequencies are independent, the total energy of a frequency
is the average number of quanta emitted at that frequency times the
energy of the frequency.

Using (\ref{P(k|nw)}), this average number of quanta of this
frequency is

\[
\overline{k}= \sum_{k=1}^{\infty} k \ P_{\Delta t}\left( k\ |\
\omega_{n_k}(\zeta_k) \right)= \left(\frac{\Delta t}{C \tau}\right)
\ \rho(\zeta_k) q(\zeta_k)^{-n_k}.
\]

Since the mean value of the number of quanta emitted at a typical
harmonic frequency $\omega_{n}(\zeta)$ is proportional to $\Delta t$
as well as $\rho(\zeta)q(\zeta)^{-n}$, the intensity of a typical
line $\omega_{n}(\zeta)$ is

\begin{equation}\label{eq. intensity}
I (\omega_{n}) = I_o\ \omega_{n}(\zeta)\ \rho(\zeta) e^{-\Lambda
\omega_{n}(\zeta)}.
\end{equation}
where $I_o$ is a constant.


\section{Temperature} In the thermal radiation from a black body
the number of quanta in a frequency is distributed by a Poisson
function, according to (\ref{P(k|nw)}). To see the consistency of
the Poisson distribution with the thermal distribution of a black
body consider the radiation from a black body at a given temperature
$T$. According to the definition of black body, the number of a
frequency $\omega$ those are emitted from within the body is
determined by the Boltzmann function, $P_T(\omega) = B\exp(- \hbar
\omega/k_B T)$, where $B$ is a normalization constant, $B=
1/\sum_{\omega}\exp(- \hbar \omega/k_B T)$, (similar to (\ref{eq.
pw|1})) and $k_B$ is Boltzman constant. The probability of $k$
quanta emissions of a specific frequency $\omega_k$ in a $j$
dimensional sequence of decays is $\left(_k^j \right)
P_T(\omega_k)^{k} \prod_{i \neq k} P_T(\omega_i)$. Summing over all
accompanying frequency except $\omega_k $, the conditional
probability is

\begin{equation}\label{eq. p_T(k|j)}
    P_T(k|\omega_k,j) =
\left(_k^j \right) B^k\exp(- k\hbar \omega_k/k_B T) (1-B\exp(- \hbar
\omega_k/k_B T))^{j-k}.
\end{equation}

Comparing this conditional probability and the one of (\ref{eq.
p(w|k,j)}), they are the same for a frequency $\omega_n$ if the
coefficients of the two exponents are equal $ \hbar /k_B T = 8 \pi
GM / c^3$. From this analogy between a black body radiation and a
black hole, one may conclude the radiation is indeed thermal and the
temperature associated to the black hole is

\begin{equation}\label{eq. T = }
T := \frac{\hbar c^3}{8 \pi G M k_B}\
\end{equation}

This coincides with the classical definition of black hole
temperature and simply performs that a radiating black hole is hot.


\section{Width of lines}

Having the above information, specially the probabilities (\ref{eq.
prob(general freq | 1)}) and (\ref{eq. pw|1}), the mean value of
emissive frequencies is easily evaluated,

\beq \langle \omega \rangle = \frac{\eta \gamma \chi \omega_o}{C},
\eeq where $\eta:=\sum_{\mathrm{all\ }\zeta}
\frac{q(\zeta)}{(q(\zeta)-1)^2}$.  A curious reader is encouraged to
follow up the easy calculation in Appendix (\ref{app5}).

By the use of the probability of decays in (\ref{eq. P(j)=}), the
mean value of the dimension of the emission sequences $j$ is $\Delta
t / \tau$. Thus, the mean decrease of the mass of black hole during
the course of $\Delta t$ is

\beq \label{eq. delta M}
 \frac{\Delta \overline{M}}{\Delta t} = -
 \frac{\hbar \langle\omega\rangle}{ c^2\tau}.
\eeq

On the other hand, the Stefan-Boltzmann law of black-body radiation
from a black hole of horizon area $A$ and surface temperature
(\ref{eq. T = }) indicates that the radiance rate from the black
hole is

\beq \label{eq. deltaM2} \frac{\Delta \overline{M}}{\Delta t} = -
\frac{\hbar c^4}{ 15360 \pi \ G^2 M^2}. \eeq

Comparing these two radiance rates of (\ref{eq. deltaM2}) and
(\ref{eq. delta M}) we can evaluate $\tau$,
\begin{equation}
\label{eq. tau = } \tau  = \frac{1920 \pi \ \eta \gamma \chi }{ C
\omega_o},
\end{equation}

By definition in (\ref{eq. P(j)=})  $\tau$ is the survival time
scale of the black hole from decaying. On average the time elapsed
before a decay is

\beq \label{eq. time average} \bar{t}=\int_{t=0}^{\infty}
tP_t(j=1)dt = 2 \tau, \eeq

The uncertainty of the elapsing time before a decay is

\beq \label{eq. time uncertainty} (\Delta t)^2 =
\int_{t=0}^{\infty}(t-\tau)^2 P_t(j=1)dt = 3\tau^2. \eeq

Due to the uncertainty principle $\Delta E \Delta t \geq \hbar/2$
and the definition of the frequency by energy, $E=\hbar \omega$, the
uncertainty of the frequency turns out to be $\Delta \omega \simeq
1/
 \tau$. Therefore, the width of emission
frequencies is proportional to $W=1/\tau$,

\begin{equation}\label{eq. width}
    W = \left(\frac{C}{1920 \pi \ \eta  \gamma \chi}\right)\ \omega_o,
\end{equation}

To estimate the order of it, let us substitute the numerical values
that are provided after $\gamma=\ln 3/\pi\sqrt{2}$. We apply the
values of $C$ and $\eta$ from the Appendices (\ref{app4}) and
(\ref{app5}).  In $SU(2)$ representation, where $\chi=0.5$,
$\eta=9.01$, and $C=2$, the ratio becomes $W=0.00029 \omega_o$. In
$SO(3)$ group representation, $\chi=\sqrt{2}$, $\eta=4.7$, and
$C=0.93$ the ratio turns out to be $W=0.00009 \omega_o$. The order
of lines width ratio is a few thousandth of the gap between the
lines, thus the spectral lines are reasonably narrow.


\section{The spectrum}

In this section, the spectrum is reviewed.

Comparing the intensities corresponding to the frequencies
$\omega_n(\zeta) = n \varpi(\zeta)$ and $\omega_m(\zeta') = m
\varpi(\zeta')$, depending on whether the generations are the same
or not, there exist two cases:

($i$) In a generation, $\zeta = \zeta'$, the relative intensity of
two harmonic frequencies is

\begin{equation}\label{eq. In/Im}
\frac{I_n}{I_m} = \frac{n}{m} q(\zeta)^{m-n}
\end{equation}

($ii$) In different generations, $\zeta \neq \zeta'$, the relative
intensities of the two modes $\omega_n\left(\zeta\right)$ and
$\omega_{n'}\left(\zeta'\right)$  is

\begin{equation}\label{eq. In/Im'}
\frac{I_n(\zeta)}{I_{n'}(\zeta')} = \frac{n}{n'}\ e^{-\Lambda
\left[n\varpi\left(\zeta\right)-n'\varpi\left(\zeta'\right)\right]}.
\end{equation}

\begin{figure}[h]
  \includegraphics[width=7cm]{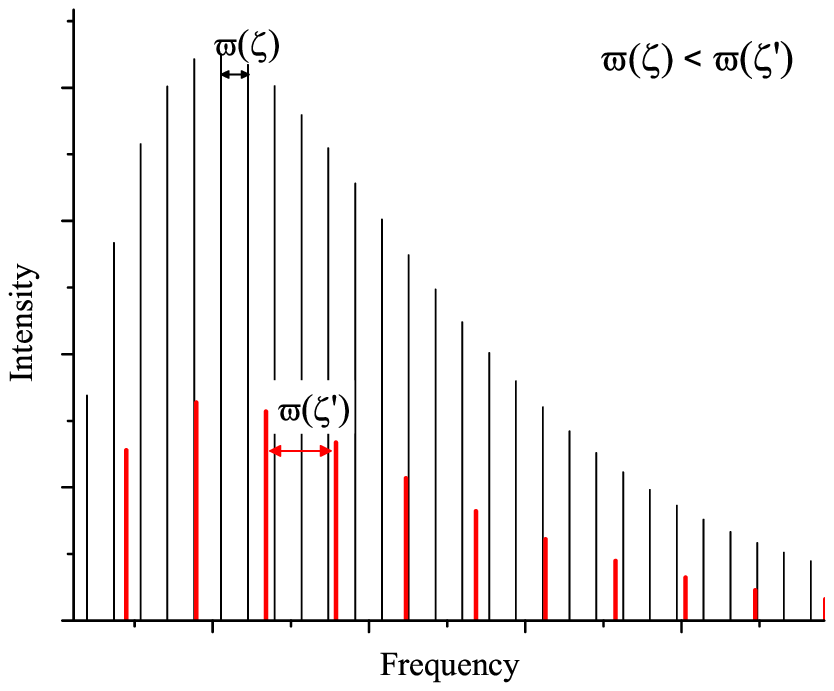}\\
  \caption{The intensities of harmonic frequencies of two generations $\zeta$ and $\zeta'$} subject to the condition $\varpi(\zeta) < \varpi(\zeta')$.\label{fig. lines}
\end{figure}

Graphically, in Fig. (\ref{fig. lines}) the intensities of harmonic
frequencies corresponding to two different generations are shown in
two different colors. The spectrum of harmonic frequencies
corresponding to the fundamental frequency $\varpi(\zeta)$ is in
black and the ones corresponding to $\varpi(\zeta')$ (for
$\omega(\zeta') > \omega(\zeta)$) is in red (the thicker set of bar
lines). The envelop of each generation matches with the one of
Hawking and Bekenstein semiclassical result.

Let us explain the disordered intensities by the following example.
Consider three  {\it consecutive} harmonic frequency modes
$\omega_1, \omega_2$ and $\omega_3$ where $\omega_i = n_i
\varpi(\zeta_i)$ for $i=1,2, 3$ and $\omega_1<\omega_2<\omega_3$.
Since $n_1, n_2$, and $n_3$ are arbitrary integers in general, let
us assume that the fundamental frequencies $\varpi_1$ and
$\varpi_3$, associated to the frequencies $\omega_1$ and $\omega_3$
respectively, are equal and the double of the fundamental frequency
$\varpi_2$ associated to $\omega_2$; ($\varpi_1 =\varpi_3$ and
$\varpi_1 = 2\varpi_2$). Since there is no other line between these
three lines, $n_3 = n_1+1 $ and $n_2= 2 n_1+1$. Comparing the
intensities associated to these three lines from (\ref{eq. In/Im}),
it turns out that the intensity of $\omega_2$ is doubled, thus the
middle line is  {\it much} brighter than the two nearby ones.

\begin{figure}
\begin{center}
\includegraphics[width=12cm]{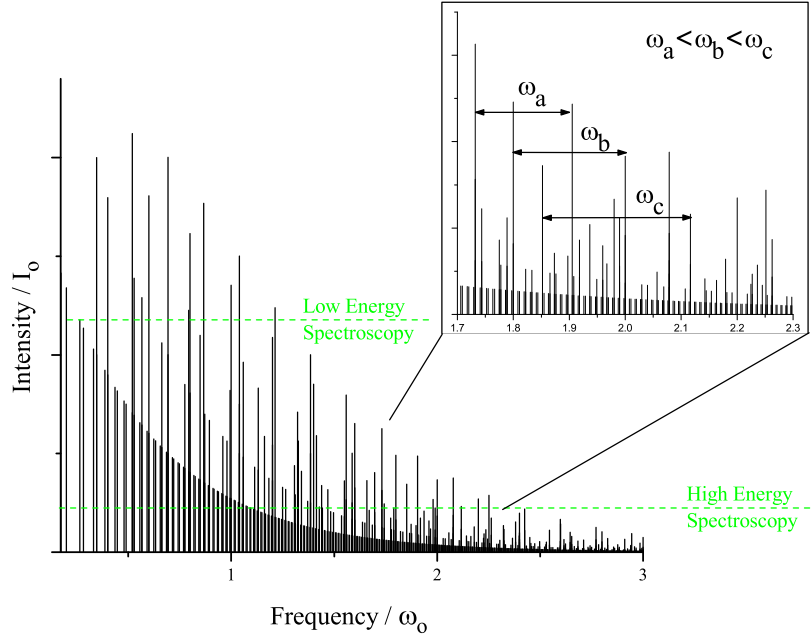}
\end{center}
\caption{ The radiation spectrum of a loop quantum black hole.}
\label{cap. Intensities}
\end{figure}

Figure (\ref{cap. Intensities}) shows the intensities corresponding
to the harmonic frequencies up to the maximum $3 \omega_o$. It is
easy to see only a countable number of the most bright line exist in
any interval of the order of $\omega_o$; those which belong to the
first few generations. In fact the intensity formula (\ref{eq.
In/Im}) shows the intensities corresponding to the third generation
on are highly suppressed relative to the first two generations.

Let us recall that in addition to these lines, there are
non-harmonic lines too but since their intensities are extremely
suppressed, they do not blend the discreteness of the most bright
lines.

The maximum intensities in the spectrum belong to the frequencies of
the condition $\omega_{\mathrm{peak}} \sim 1/\Lambda=c^3/8 \pi GM =
\omega_o/ \pi$. Therefore, the most bright lines are of the
harmonics of integer valued number $n_{\mathrm{peak}} \sim (\gamma
\pi \chi \sqrt{\zeta})^{-1}$. Among all of the parameter, only the
Barbero-Immirzi parameter is not certainly known and the dependency
of the peak to the parameter is remarkable for the purpose of a
possible way to determining it.


\section{Discussion}

The discreteness of area eigenvalues comes about the canonical
quantization of 3-geometry because it is supposed that geometry has
a distributional character with 1-dimensional excitations. Having
this, the quantum geometrical operators are constructed by the
canonical variables of loop quantum geometry. Among them, the area
operator is the one whose corresponding eigenvalues are completely
known.

In part (I) it was demonstrated that the area eigenvalues exhibits
an unexpected symmetry. In fact, the spectrum of the numbers can be
split into equidistant sequences of numbers. Each one of these
evenly spaced sets of numbers is called a `generation.' Each
generation possesses an individual gap between levels, by which it
is identified. The gap is proportional to the square root of a
square-free number in $SO(3)$ representation, or the discriminant of
a positive definite quadratic form in $SU(2)$ representation.

Consequently, the eigenvalues of area operator, instead of being
labeled by three free numbers $j_u$, $j_d$ and $j_{u+d}$, can be
performed by fewer numbers; which can be the representative that
specifies the generation and the level within a generation.

The relation between area and mass of a black hole (valid only on a
black hole horizon), introduces quanta of energy by the use of the
`area' states of horizon. Having the symmetry of the quanta of
horizon area, two different types of area transitions are possible:
the transition either ($i$) between area levels within a generation
(the o-called `generational transitions') or ($ii$) between the area
levels of different generations (the so-called `inter-generational
transitions'). One of the immediate consequences of this symmetry is
there appears a discrimination between these two types of
transitions. Those quanta emitting from generational transitions can
be reproduced in many copies from many levels of a generation.
However, there exists only one copy of each inter-generational
transitions. This leads to a discrimination in the population of
generational transitions motivated by the quantum amplification
effect.

In Part (II) the intensity of radiation for any frequency was worked
out. It was illustrated that a black hole radiates a continuous
spectrum of frequencies. The spectrum of the quanta frequencies
ranges from zero to a maximum. Nonetheless, there exist some
spectral lines which take {\it additional} intensities due to the
quantum amplification effect. This `amplification' is a features of
loop quantization of area. Following this, black hole radiation is
dominated by the amplified area fluctuations and some discrete
bright lines appear.

The smaller a fundamental frequency is, the more bright the
harmonics are. Due to the $\gamma$-dependency of the intensity
function and according to figure (\ref{cap. Intensities}) the most
bright lines in various energy scales of the spectrum belong to the
first (or a few of the first) generation. Since the spectral lines
are sufficiently narrow and apart from each other, they unlikely
blend. In fact, the width of the lines are expected to be of the
order of a thousandth of the frequency scale factor $\omega_o$,
while the gap between intensity peaks are of the order of this
factor. Thus it is expected that such a quantum black hole radiates
in a {\it visually discrete} pattern.

The precise spectroscopy depends on the exact value of the
Barbero-Immirzi parameter as well as the group representation of
spin network states.

Among the possible predictions of a canonically quantized black
holes there are some features: 1) the radiation is effectively is
visually discrete to observation, and 2) the intensities of
consecutive lines are not orderly distributed.

\begin{figure}
\begin{center}
\includegraphics[width=6cm]{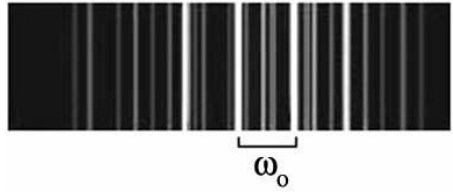}
\end{center}
\caption{A typical spectrum of a canonically quantized black hole
radiation for the low energy spectroscopy of Fig (\ref{cap.
Intensities}).} \label{fig.spectrum}
\end{figure}

Figure (\ref{fig.spectrum}) is a typical expected radiance spectrum
of a canonically quantized black hole is generated in low energy
spectroscopy of figure (\ref{cap. Intensities}). If the actual
spectrum of black radiation is effectively discrete, the detection
of a few of the most bright lines will be adequate to justify
experimentally this prediction. The most bright lines in the
spectrum belong to the first generation and the gap is of the order
of the frequency scale $\omega_o \sim 10^{16}/ M_{\mathrm{(kg)}}$
(eV).

Recently, a number of efforts have been put on the discovery of the
radiance patterns of different types of black holes, the primordial
holes \cite{{Cline:2001tq},{Frampton}} and the one of higher
dimensions \cite{Barcelo:2001ca}. One way to detect primordial black
holes is by their Hawking radiation. The prediction of a canonically
quantum black hole is also amenable to experimental check if the
primordial black holes are founded. For instance, if the primordial
black holes constitute an essential part of dark matter in the
galactic distant, the observation of a few of their most bright
radiance lines can be within the modern sensitivity and can be
possibly distinguished from the radiation of other objects.

It should also be noted that this radiation, if associated to the
primordial black holes, is far beyond the Trans-Planckian problem of
inflationary cosmology. The Trans-Planckian problem refers to the
derivation of physical quantities from quantum field theory beyond
the Planck scale. However, the proposed spectroscopy is based on a
version of quantum gravity in which the difficulties within the
semi–classical approximation does not exist.

Among important questions that are asked about the developments of
area operator and black hole physics, there remain some important
questions.

From experimental point view,

\begin{itemize}
  \item One of the most incredibly important questions is that how far are we from detecting this spectrum?
  \item Considering a tiny percentage of dark matter obtaining from primordial black holes, is it possible to verify the spectrum as an alternative instead of gamma-ray busters?
  \item In the case of a rotating black holes, how the spectral lines are shifted or
  widen?
\end{itemize}

From numerical point of view,

\begin{itemize}
  \item What is the correlation between the Barbero-Immirzi parameter ($\gamma$) and the visual spectrum at different energy spectroscopies? In other words, considering the $\gamma$-dependency of the intensities, what are the visual frequencies in a low energy spectroscopy?
 \end{itemize}

From theoretical point of view,

\begin{itemize}
  \item Under imposing what conditions the quantum dynamics of black hole spin network states are identified?
  \item What is the Planck scale corrections to the entropy of such a quantum black hole?
\end{itemize}

It will be also interesting to see if a similar pattern can be
illustrated for near extremal black holes in supergravity and string
theories.

\section{Acknowledgement}
Helpful discussions with Lee Smolin are acknowledged.

\part{Appendix}

\appendix


\section{Area spectrum in $SO(3)$ version} \label{app1}

\begin{description}
  \item[Theorem 1:] The set of numbers evaluated by the generating formula $\frac{1}{2} [2a(a+1) + 2b(b+1) - c(c+1)]$, where $a$, $b$, and $c$ are positive integers and $c \in \{|a-b|, \cdots, a+b-1, a+b \}$,
  is reduced into the whole $\mathbb{Z}^+$, modulo rematching.
  \item[\it Proof:] Suppose $a \geq b$.  Let us consider the two independent numbers are $a = b + n$, where $n$ is a positive integer.
  The subset $c = a + b$ is generated by the formula $n(n + 1)/2 +
  b$. The first term is called Triangular numbers and are integers. The second term is independent from the first term and can be any positive
  integer. This subset generates all positive integers and since other
  subsets generate integers, all of them fit into the whole positive integer sets
  $\mathbb{Z}^+$. This set is a reduced set of the original one subject to identifying all repeated numbers.
\end{description}

The fundamental theorem of arithmetic states that every positive
integer (except the number 1) can be represented in exactly one way
as a product of one or more primes, apart from rearrangement. This
theorem is also called `the unique factorization theorem'.  Thus,
prime numbers are the `basic building blocks' of the natural
numbers.\footnote{In number theory, the prime factors of a number
are considered as indistinguishable building blocks of numbers and
thus the ordering of numbers does not matter.} Decomposing any
natural number into its prime numbers, the primes are either
repeated or not. Collecting the natural numbers whose prime factors
are not repetitive the square-free sequence of numbers are produced.

 {$\blacktriangleright $}
\textbf{Definition of Square-free Numbers:} an integer number is
said to be square-free, if its prime decomposition contains no
repeated factors. For example, $30$ is square free since its prime
decomposition $2\times3\times5$ contains no repeated factors. In
this note, this sequence is indicated by the symbol $\mathbb{A}$.
Some of the known square-free numbers are given in table (\ref{tab. square-free}).
\begin{table}[h]
\begin{tabular}{|p{12cm}|}\hline\begin{center}
 \begin{tabular}{p{11cm}}
$\mathbb{A}=\{$1, 2, 3, 5, 6, 7, 10, 11, 13, 14, 15, 17, 19, 21, 22,
23, 26,
 29,
 30, 31, 33, 34, 35, 37, 38, 39, 41, 42, 43, 46, 47, 51, 53, 55, 57, 58, 59, 61, 62,
 65,
 66, 67, 69, 70, 71, 73, 74, 77, 78, 79, 82, 83, 85, 86, 87, 89, 91, 93, 94, 95, 97, 101,
 102,
 103, 105, 106, 107, 109, 110, 111, 113 , $\cdots \}$.
\end{tabular}
\end{center}\\ \\ \hline
\end{tabular}
\caption{Square-free numbers (Sloane's A005117)}
\label{tab. square-free}
\end{table}

There is no known polynomial algorithm for recognizing square-free,
\cite{Niven1966}. {$\blacktriangleleft$}

A natural number is the multiplication of a square-free number and a
square number.

\begin{description}
  \item[Theorem:] The natural numbers can be rewritten as a mixture of square
generations by the contribution of all square-free representatives,
\begin{equation}\label{eq. N=U_N^2} \{\mathbb{N}\} \equiv
\bigcup_{\zeta \in \mathbb{A}} \{\zeta \mathbb{N}^2\}.
\end{equation}
 \item{\it Proof:} Any natural
number can be written in terms of its prime factors, say
$p_1^{n_1}\times p_2^{n_2}\cdots \times p_i^{n_i}$, where $p_1,$
$p_2$, $\cdots$, and $p_i$ are different prime numbers and the
exponents $n_1$, $n_2$, $\cdots$, $n_i$ are positive integers. These
exponents are either even or odd numbers. In the most general case
all of the exponents are different odd numbers, $n_i=2m_i +1$.
Therefore, the above-mentioned number can be rewritten in the form
$(p_1 \times p_2 \cdots \times p_i) \times (p_1^{m_1}\times
p_2^{m_2}\cdots \times p_i^{m_i})^2$. Due to the assumption that the
prime number $p$'s are different, the first parenthesis is
equivalent to a square-free number and the second parenthesis is
nothing but $n^2$ for the natural number $n = p_1^{m_1}\times
p_2^{m_2}\cdots \times p_i^{m_i}$. Therefore, $\forall\ x \in
\mathbb{N}, \ \ \exists\  y \in \mathbb{N} \ \ \mathrm{and}\ \ a \in
\mathbb{A}, \ \ \ x \equiv a \times y^2 $.
\end{description}

In Table \ref{tab. 10squarefree}, having the first 15 square-free
numbers of $\mathbb{A}$, the corresponding elements of the square
generations $\zeta \mathbb{N}^2$ are tabulated up to the first five
elements.

\begin{table}[h]
  \centering
  \begin{tabular}{|c|c|c|c|c|c|c|c|c|c|c|c|c|c|c|c|}
  \hline & $1 \mathbb{N}^2$ & $2 \mathbb{N}^2$ & $3 \mathbb{N}^2$ & $5 \mathbb{N}^2$ &
  $6 \mathbb{N}^2$ &
   $7 \mathbb{N}^2$ & $10 \mathbb{N}^2$ &  $11 \mathbb{N}^2$ &  $13 \mathbb{N}^2$ & $14 \mathbb{N}^2$ & $15 \mathbb{N}^2$ & $17 \mathbb{N}^2$ & $19 \mathbb{N}^2$ & $21 \mathbb{N}^2$\\
   \hline
   m=1\footnote{This row shows the representatives $\zeta$.}&1&2&3&5&6&7&10&11&13&14&15&17&19&21 \\
\hline

   m=2&4&8&12&20&24&28&40&44&52&56 &60 &68 &76 &84 \\ \hline

   m=3&9&18&27&45&54&63&90&99&117&126 &135 &153 &171 & 189\\ \hline

   m=4&16&32&48&80&96&112&160&176&208&224 &240 &272 &304 & 336\\ \hline

   m=5&25&50&75&125&150&175&250&275&325&350 &375 &425 &475 & 525\\ \hline






   \end{tabular}
\caption{The first fifteen elements of some $SO(3)$ based
generations $\zeta \mathbb{N}^2$.}\label{tab. 10squarefree}
\end{table}

A column in the Table (\ref{tab. 10squarefree}) indicates the
elements of a square generation and consists of all natural number
up to 21. By extending this Table, the consistency of the elements
with natural numbers can be verified up to any order. There is no
common element in different square generations,
\begin{equation}\label{eq. uniqunessofsquarefree}
\forall\  \zeta_1, \zeta_2 \in \mathbb{A}, \ \ \mathrm{if}\ \zeta_1
\neq \zeta_2,\ \ \ \{\zeta_1 \mathbb{N}^2\} \cap \{\zeta_2
\mathbb{N}^2\} = \emptyset.
\end{equation}

Consider the sequence of numbers that contain the same square-free
$\zeta$ factor multiplied by all square numbers, $\zeta
\mathbb{N}^2$. Taking square root from the elements of the square
generation the  {\it equidistant} sequence $\sqrt{\zeta} \mathbb{N}$
is produced.


\section{Area spectrum in $SU(2)$ version} \label{app2}

\begin{description}
  \item[Theorem:] The set of numbers evaluated by the generating formula $4 [2a(a+1) + 2b(b+1) - c(c+1)]$, where $a$ and $b$ are positive integer or half-integer of $\frac{1}{2}\mathbb{Z}$ and $c \in \{|a-b|, \cdots, a+b-1, a+b \}$,
  is reduced into the congruent number unto 0 and 3 mode 4 modulo the degeneracies.
  \item[\it Proof:] Suppose $a \geq b$.  Let us consider the two independent numbers are $a = b + n$, where $n \in \frac{1}{2}\mathbb{Z}$.
  The subset $c = a + b$ is generated by the formula $4n(n + 1) + 8b$. Let us consider $n = N/2$ and $b = B/2$ where $N$ and $B$ are independent natural numbers. Substituting them in the formula
  it becomes $N(N+2) + 4B$. The first term is the mixture of congruent numbers unto 0 or 3 mod 4. The second term is the congruent numbers unto 0 mod 4. In other words, a number that is generated by $N(N+2)$ is either $4m+3$ or $4m$ for some integers $m$. This is not changed
  when the term $4B$ is added to the numbers. Let us fix $N$ unto either 0 or 1. The whole sequence of congruent numbers unto 0 and 3 mod 4 are obviously generated from $4B+3$ and $4B$ for any integer $B$. All other numbers fit to the whole sequence if one identifies all
degeneracies.
\end{description}

Evaluating $4(m_{j_u, j_d,j_{u+d}})^2$ in $SU(2)$ group
representation, the Skew Amenable numbers are produced.

 {$\blacktriangleright$} {\bf
Definition of Skew Amenable numbers}: in a simple definition these
numbers are the page numbers of a book that is printed out only at
the pages that come after each two leaves of sheets by face. This
can be interpreted in a mathematical language as the congruent
numbers to either 0 or 3 in mod 4. \footnote{ There is also another
definition that a number $n$ is skew amenable if there exist a set
of integers $\{m_i\}$ satisfying the relations: $n = \sum_{i=1}^n
m_i = - \prod_{i=1}^n m_i$, \cite{Barger}. For instance, the number
8 is a skew amenable because is can be decomposed into an 8 term sum
as well as the negated product of exactly the same numbers:
$8=1+1+1+1+1+1-2+4=-(1\times1\times1\times1\times1\times1\times(-2)\times4)$.
Another example is 3, which satisfied the condition:
$3=1+3-1=-(1\times 3 \times (-1))$.}

The Skew Amenable numbers smaller than 200 are  3, 4, 7, 8, 11, 12,
15, 16, 19, 20, 23, 24, 27, 28, 31, 32, 35, 36, 39, 40, 43, 44, 47,
48, 51, 52, 55, 56, 59, 60, 63, 64, 67, 68, 71, 72, 75, 76, 79, 80,
83, 84, 87, 88, 91, 92, 95, 96, 99, 100, 103, 104, 107, 108, 111,
112, 115, 116, 119, 120, 123, 124, 127, 128, 131, 132, 135, 136,
139, 140, 143, 144, 147, 148, 151, 152, 155, 156, 159, 160, 163,
164, 167, 168, 171, 172, 175, 176, 179, 180, 183, 184, 187, 188,
191, 192, 195, 196, 199, 200, etc. (Sloane's A014601)\footnote{
http://www.research.att.com/~njas/sequences/A014601}
{$\blacktriangleleft$}

It is known in number theory that any square integer number is the
congruent to either 0 or 1 mod 4. On the other hand, there is a
theorem that in the existence of two equalities $x_1\equiv x_2$ (mod
$m$) and $x_3 \equiv x_4$ (mod $m$) of the same modular $m$, it can
be it is easy to verify that $x_1 \times x_3 \equiv x_2 \times x_4$
(mod $m$). Accordingly, having one of the two equality as $x_1=0$ or
3 (mod4) for a Skew Amenable number $x_1$, and the equality $x_2=0$
or 1 (mod4) for a square number $x_2$, the multiplication of these
two produces a Skew Amenable number $x_1 \times x_2=0$ or 3 (mod4)
is generated. In other words, multiplying the complete set of Skew
Amenable numbers and the complete set of square numbers, the product
a subset of the Skew Amenable numbers is generated.

Having this fact in mind, for any random Skew Amenable number $b'$
there exists a corresponding Skew Amenable number $b$ that satisfies
the equality $b'=b \times n^2$ for $n \in \mathbb{N}$. In fact, the
set of numbers $b$ for all $n \in \mathbb{N}$ is a subset of Skew
Amenable numbers. We represent this subset by the symbol
$\mathbb{B}$. Now the question is: what is $\mathbb{B}$? To answer
the questions, the Skew Amenable number can be generated and the
elements of the set $\mathbb{B}$ are identified individually. The
result is in the table (\ref{tab. B}).
\begin{table}[h]
  \centering
\begin{tabular}{|p{12cm}|}\hline
\begin{center}
\begin{tabular}{p{11cm}}
$\mathbb{B}=\{$3, 4, 7, 8, 11, 15, 19, 20, 23, 24, 31, 35, 39, 40,
43, 47, 51, 52, 55, 56, 59, 67, 68, 71, 79, 83, 84, 87, 88, 91, 95,
103, 104, 107, 111, 115, 116, 119, 120, 123, 127, 131, 132, 136,
139, 143, 148, 151, 152, 155, 159, 163, 164, 167, 168, 179, 183,
 184, 187, 191, $\cdots \}$. \\
\end{tabular}
\end{center} \\ \\ \hline
\end{tabular}
\caption{The Discriminants of the Positive Definite Quadratic Forms
(Sloane's A003657)}\label{tab.
B}
\end{table}

The negative of this sequence of numbers coincide with a well-known
sequence of discriminants of  {\it the Positive Definite Quadratic
Forms}, \cite{Buell}. The definition of the positive definite
quadratic forms is explained in the Appendix (\ref{appendix}) of
this note.

Consequently, an Skew Amenable sequence of number, which is
generated from the evaluation of $4(m_{j_u, j_d,j_{u+d}})^2$ in
$SU(2)$ representation, can be rewritten as an element of the set
$\mathbb{B}$ multiplied by an integer squared. In other words, the
elements of the set $4(m_{j_u, j_d,j_{u+d}})^2$ can be represented
as square generations with representative elements of the set
$\mathbb{B}$. In table \ref{tab. Skewtable} sixteen elements of the
square generations whose representatives are the first five elements
of the set $\mathbb{B}$, is tabulated.
\begin{table}[h]
  \centering
  \begin{tabular}{|c|c|c|c|c|c|c|c|c|c|c|c|c|c|c|c|c|}
  \hline & $3\mathbb{N}^2$ & $4\mathbb{N}^2$ & $7\mathbb{N}^2$ & $8\mathbb{N}^2$ &
  $11\mathbb{N}^2$ &
   $15\mathbb{N}^2$ & $19\mathbb{N}^2$ &  $20\mathbb{N}^2$ &  $23\mathbb{N}^2$ & $24\mathbb{N}^2$ & $31\mathbb{N}^2$ & $35\mathbb{N}^2$ & $39\mathbb{N}^2$ & $40\mathbb{N}^2$ &$43\mathbb{N}^2$ \\
   \hline

   m=1\footnote{This row shows the representatives $\zeta$.}&3&4&7&8&11&15&19&20&23&24&31&35&39&40&43 \\
\hline

   m=2&12&16&28&32&44&60&76&80&92&96&124&140&156&160 &172\\ \hline

   m=3&27&36&63&72&99&135&171&180&207&216&279&315&351&360& 378\\ \hline

   m=4&48&64&112&176&240&304&320&368&384&384&496&560&624&640 &688\\ \hline

   m=5&75&100&175&200&275&375&475&500&575&600&775&875&975&1000&1075 \\ \hline






   \end{tabular}
\caption{The first sixteen elements of some $SU(2)$ based
generations $\zeta \mathbb{N}^2$.}\label{tab. Skewtable}
\end{table}

In the table (\ref{tab. Skewtable}), all elements of the Skew
Amenable numbers up to 44 are present and any extension of the table
will produce all of the others up to any order.

There is no common element in different square generations,
\begin{equation}\label{eq. uniqunessofsquarefreesu2}
\forall\  \zeta_1, \zeta_2 \in \mathbb{B}, \ \ \mathrm{if}\ \zeta_1
\neq \zeta_2,\ \ \ \{\zeta_1 \mathbb{N}^2\} \cap \{\zeta_2
\mathbb{N}^2\} = \emptyset.
\end{equation}


\section{Positive definite quadratic forms}\label{appendix}

A quadratic form is a two-variable integer-valued function $f(x,y) =
ax^2 + bxy + cy^2$, with $a, b, c \in \mathbb{Z}$. This form
`primitive' if $a, b, c$ are relatively `prime'. The `discriminant'
of this form is defined as $\Delta := b^2 - 4ac$.

Substituting integers values in two variables $x$ and $y$
respectively, the form is evaluated by an integer $m$, $f(x_0, y_0)
= m$. This problem can be restated as follows, the `representation'
of the form $f(a, b, c) = m$ is elements of the solution space of
the equation $(x_0, y_0)$. A representation is called `primitive' if
gcd($x_0, y_0$)=1.

Given a form $f$, the transformation $x = \alpha x' + \beta y' $ and
$y = \gamma x' + \delta y '$ transforms $f$ into $f'$. The new form
$f'$ remains as an integer if and only if $\alpha\delta -
\beta\gamma = \pm 1$. The interesting fact is that in such a
transformation that preserves the integer character of the form the
discriminant $\Delta$ remains  {\it invariant}.

If $f$ is a form of integer $m$, we can rewrite the definition of a
form as $4 am = (2ax + by)^2 - \Delta y^2 $. In the case that
$\Delta$ is a perfect square number the right hand side is written
in the form $(2ax + by + y \sqrt{\Delta})(2ax + by - y
\sqrt{\Delta})$. In this case the different representations of
solutions are  {\it degenerate} and thus indistinguishable. These
forms are of this note.

In the case of $\Delta<0$, it is clear from $4 am = (2ax + by)^2 -
\Delta y^2 $ that for any representation $(x, y)$, $m$ and $a$ (and
$c$) are of the same sign. A forms whose corresponding discriminant
is negative is called a `definite form' and if $m$ is positive, it
is called a `positive definite form'.\footnote{ In the case that
$\Delta>0$, $a$ and $c$ are of opposite signs and thus both positive
an negative integers $m$ may be represented on $f$. This case is
called indefinite form and is not of our interest of study.}

By substituting the positive $a$ and $c$ in $f(x,y) = ax^2 + bxy +
cy^2$, if the form for any representation $(x, y)$ become positive
and the discriminant becomes negative, the form is a definite
positive quadratic form. Evaluating the negated values of the
discriminants of such forms produces the following sequence of
numbers 3, 4, 7, 8, 11, 15, 19, 20, 23, 24, 31, 35, 39, 40, 43, 47,
51, 52, 55, 56, 59, 67, 68, 71, 79, 83, 84, 87, 88, 91, 95, 103,
104, 107, 111, 115, 116, 119, 120, 123, 127, 131, 132, 136, 139,
143, 148, 151, 152, 155, 159, 163, 164, 167, 168, 179, 183,
 184, 187, 191, $\cdots$, which is the same elements of the set
$\mathbb{B}$ in table (\ref{tab. B}).

For instance, $(a=1, b=1, c=1)$ defines a positive definite form
whose corresponding discriminant is -3 and its negated value is the
first element of $\mathbb{B}$. The second element is produced after
$(a=1, b=0, c=0)$.

To verify that this sequence is the congruent to either 0 or 3 mod4,
it is enough check the consistency of the negated discriminant $-
\Delta = 4ac - b^2$ with this congruent. It is clear that $b^2$ is
congruent to 0 or 1 (mod4). Also, $4ac$ is the congruent to 0 mod4.
Therefore, $-\Delta$ is congruent to 0 or 3 (mod4).



\section{The normalization coefficient $C$} \label{app4}

To calculate this coefficient, it should be noticed the spectrum of
non-harmonics is almost continuous except at zero and the harmonics,
$\omega' \in \mathbb{R}^+ - \{0\} - \{ \mathrm{harmonics}\}$. These
frequencies are all uniformly weighted by $\rho = 1/N_o$.  Since the
population of harmonics is much more than the non-harmonics, we can
approximate the population of a harmonics to be $N-1$ instead of $N$
and add the one copy of each harmonic into the above mentioned set
of frequencies in order to fill the gaps. Doing so, the equally
weighted set of frequencies $\omega' \in \mathbb{R}^+ - \{0\}$ is
provided. By the use of (\ref{eq. prob(general freq | 1)}) and
(\ref{eq. pw|1}), in the classical limits ($N_o \gg 1$), the
normalization coefficient reads
\begin{eqnarray} \label{eq. C approx formula}
C &=& \lim_{N \rightarrow \infty}  \sum_{\mathrm{all}\ \zeta}\
\sum_{n=1}^{N} \frac{N-1}{N}\rho(\zeta)\ q(\zeta)^{-n} +
 \int  \frac{1}{N_o}\
e^{-\Lambda \omega'} \Lambda d\omega' \nonumber\\
&\simeq& \sum_{\mathrm{all}\ \zeta}\ \sum_{n=1}^{\infty}
\rho(\zeta)\ q(\zeta)^{ - n}.
\end{eqnarray}

By the use of the algebraic relation $\sum_{n=1}^{\infty} x^{-n} =
1/ (x-1)$, the internal sum in $C$ is summarized. Consequently, the
normalization coefficient becomes
\begin{equation}\label{eq. C}
    C = \sum_{\mathrm{all}\ \zeta} \ \frac{\rho(\zeta)}{q(\zeta)-1}
\end{equation}

It is useful to check the finiteness of the normalization
coefficient $C$.
\\

\begin{quote}
 \textbf{
The finiteness of C:} for the purpose of simplicity the definition
$q(\zeta) = \exp( \pi \gamma \chi \sqrt{\zeta})$ can be rewritten
$q(\zeta):= h^{\sqrt{\zeta}}$, where $h:= \exp( \pi \gamma \chi)$ is
in both groups greater than 1.

The Cauchy root method of convergence test is such that for series
like $\sum_n a_n$, if the value of $\lim_{n\rightarrow \infty}
|a_{n}|^{1/n}$ is smaller than one, the series converges. In the
series sum of $C$, this condition reads
\begin{equation}\label{eq convergence}
\lim_{\sqrt{\zeta}\rightarrow \infty} \left| \frac{1}{\sqrt{\zeta}}\
\frac{1}{h^{\sqrt{\zeta}}-1}\right|^{\frac{1}{\sqrt{\zeta}}} \sim
\lim_{\sqrt{\zeta}\rightarrow \infty} \left(
\frac{h^{-\sqrt{\zeta}}}{\sqrt{\zeta}}\right)^{\frac{1}{\sqrt{\zeta}}}
= \frac{1}{h} < 1.
\end{equation}

Therefore $C$ is a finite number.
\\
\end{quote}

Numerical work can estimate the range of $C$. Substituting $\rho$
from (\ref{eq. rho}) and Barbero-Immirzi parameter from
\cite{Ansari:2006cx}$, \gamma = \ln 3/\pi \sqrt{2}$, in $q(\zeta)$
the coefficient $C$ is simplified to $\sum_{\zeta}\zeta^{-1/2}(3^{
\chi \sqrt{\zeta/2}}-1)$. In the $SU(2)$ representation, where
$\zeta \in \mathbb{B}$ and $\chi=1/2$, turn out to be  $C=2.01$,
whilst in the $SO(3)$ representation group, where $\zeta \in
\mathbb{A}$ and $\chi=\sqrt{2}$,  it is $C= 0.93$.


\section{Average of frequency $\langle\omega\rangle$} \label{app5}

Since the probabilities of (\ref{eq. prob(general freq | 1)}) and
(\ref{eq. pw|1}) are normalized, the mean value of the emitting
frequencies is

\begin{equation*}
    \langle\omega\rangle := \sum_{\zeta}\sum_{n =1}^{\infty}
\omega_n(\zeta)\ P_{\Delta t}(\{\omega_n(\zeta)\}\ |\ 1) +
\left(\frac{1}{N_o}\right) \int_{0}^{\infty}\omega e^{-\Lambda
\omega} \Lambda d\omega.
\end{equation*}

The second term for a classical black hole is negligible comparing
to the first term.  Using (\ref{eq. pw|1}) and the algebraic formula
$\sum_{n=1}^{\infty} n x^n = x/(1-x)^2$, where $x<1$, the mean value
of the frequency of a generation $\zeta$ turns out to be $(1/C)\
\varpi(\zeta) \ \rho(\zeta) q(\zeta)/\left(q(\zeta)-1\right)^2$ and
therefore the mean value of all frequencies becomes

\begin{equation}\label{eq. w bar1}
\langle\omega\rangle \sim \frac{1}{C} \sum_{\zeta} \varpi(\zeta) \
\frac{\rho(\zeta)\ q\left(\zeta\right)}{\left(q(\zeta)-1\right)^2}
\end{equation}

Using (\ref{eq. w^eta = }) and (\ref{eq. rho}), we can rewrite the
equation in the from

\beq \langle\omega\rangle \sim \frac{\omega_o \gamma \chi}{C} \
\sum_{\zeta} q\left(\zeta\right) / \left(q(\zeta)-1\right)^2. \eeq

\begin{quote}

{\bf Convergence of $\langle\omega\rangle$:} To check the
convergence of the sequence $\sum_{\zeta} q(\zeta) / (q(\zeta)-1)^2$
via the Cauchy's convergence test, since the real free index in the
function $q(\zeta)$ is $\sqrt{\zeta}$ the test function
$\lim_{\sqrt{\zeta}\rightarrow \infty}\ |q(\zeta) /
(q(\zeta)-1)^2|^{1/\sqrt{\zeta}}$ should be considered. By the use
of the definition of $q(\zeta) := h^{\sqrt{\zeta}}$, where
$h:=\exp(\pi \chi \gamma)>1$,

\begin{equation}\label{eq convergence}
\lim_{\sqrt{\zeta}\rightarrow \infty} \left| \frac{1}{\sqrt{\zeta}}\
\frac{h^{\sqrt{\zeta}}}{(h^{\sqrt{\zeta}}-1)^2}\right|^{\frac{1}{\sqrt{\zeta}}}
\sim  \frac{1}{h} < 1.
\end{equation}

This summation converges.
\end{quote}

Let us rewrite the sum by the use of the definition
$\eta:=\sum_{\zeta} q(\zeta) / (q(\zeta)-1)^2$ as $\langle \omega
\rangle = \eta \gamma \chi \omega_o /C$. Having the parameter from Appendix (\ref{app4}), and by
substituting $\gamma=\ln 3/\pi\sqrt{2}$ the coefficient $\eta$ in
the $SU(2)$ representation turns out to be $\eta=9.0$, while it is
$\eta=1.7$ in the $SO(3)$ group representation. By the use of the
numerical values of $C$, the mean value of frequency $\langle \omega
\rangle$ is either about $11 \omega_o$ in $SU(2)$ group or about $15
\omega_o$ in $SO(3)$ group.


\begin{thebibliography}{99}

\bibitem{Helfer:2003va}
  A.~D.~Helfer,
   {\it ``Do black holes radiate?,''}
  Rept.\ Prog.\ Phys.\  {\bf 66}, 943 (2003)
  [arXiv:gr-qc/0304042].

\bibitem{Wald:1999vt}
  R.~M.~Wald,
   {\it ``The thermodynamics of black holes,''}
  Living Rev.\ Rel.\  {\bf 4}, 6 (2001)
  [arXiv:gr-qc/9912119];
  T.~Padmanabhan,
   {\it ``Gravity and the thermodynamics of horizons,''}
  Phys.\ Rept.\  {\bf 406}, 49 (2005)
  [arXiv:gr-qc/0311036].

\bibitem{Bekenstein:1995ju}
  Bekenstein et. al.
   {\it ``Spectroscopy of the quantum black hole,''}
  Phys.\ Lett.\ B {\bf 360}, 7 (1995) [arXiv:gr-qc/9505012].


\bibitem{Smolin:2006pa}
  L.~Smolin,
   {\it ``Generic predictions of quantum theories of gravity,''}
  [arXiv:hep-th/0605052];
  A.~Ashtekar and J.~Lewandowski,
   {\it ``Background independent quantum gravity: A status report,''}
  Class.\ Quant.\ Grav.\  {\bf 21}, R53 (2004)
  [arXiv:gr-qc/0404018].


\bibitem{Markopoulou:2006qh}
  F.~Markopoulou,
   {\it ``Towards gravity from the quantum,''}
  arXiv:hep-th/0604120;
  J.~Ambjorn, J.~Jurkiewicz and R.~Loll,
   {\it ``Quantum gravity, or the art of building spacetime,''}
  arXiv:hep-th/0604212.

\bibitem{Smolin:1994ge}
  C.~Rovelli and L.~Smolin,
  {\it ``Discreteness of area and volume in quantum gravity,''}
  Nucl.\ Phys.\ B {\bf 442}, 593 (1995)
  [Erratum-ibid.\ B {\bf 456}, 753 (1995)]
  [arXiv:gr-qc/9411005];

  \bibitem{Ashtekar-surface}
 A. Ashtekar et. al.,
   {\it ``Quantum theory of geometry. I: Area operators,''}
  Class.\ Quant.\ Grav.\  {\bf 14}, A55 (1997).
  [arXiv:gr-qc/9602046].


\bibitem{Ashtekar:black hole}
  A.~Ashtekar, J.~C.~Baez and K.~Krasnov,
   {\it ``Quantum geometry of isolated horizons and black hole
  entropy,''}
  Adv.\ Theor.\ Math.\ Phys.\  {\bf 4}, 1 (2000)
  [arXiv:gr-qc/0005126];
  A.~Ashtekar,
   {\it ``Interface of general relativity, quantum physics and statistical
  mechanics: Some recent developments,''}
  Annalen Phys.\  {\bf 9}, 178 (2000)
  [arXiv:gr-qc/9910101];
  A.~Ashtekar, J.~Baez, A.~Corichi and K.~Krasnov,
  {\it``Quantum geometry and black hole entropy,''}
  Phys.\ Rev.\ Lett.\  {\bf 80}, 904 (1998)
  [arXiv:gr-qc/9710007];
  A.~Ashtekar and B.~Krishnan,
   {\it ``Isolated and dynamical horizons and their applications,''}
  Living Rev.\ Rel.\  {\bf 7}, 10 (2004)
  [arXiv:gr-qc/0407042].

\bibitem{Barreira:1996dt}
  Barreira, et. al.
   {\it ``Physics with nonperturbative quantum gravity: Radiation from a quantum
  black hole,''}
  Gen.\ Rel.\ Grav.\  {\bf 28}, 1293 (1996) [arXiv:gr-qc/9603064].


\bibitem{Ansari:2006cx}
  M.~Ansari,
   {\it ``Genericness of degeneracy and entropy in loop quantum gravity,''}
  arXiv:gr-qc/0603121.



\bibitem{Dreyer:2002vy}
  O.~Dreyer,
   {\it ``Quasinormal modes, the area spectrum, and black hole
  entropy,''}
  Phys.\ Rev.\ Lett.\  {\bf 90}, 081301 (2003)
  [arXiv:gr-qc/0211076];
  L.~Motl,
   {\it ``An analytical computation of asymptotic Schwarzschild quasinormal
  frequencies,''}
  Adv.\ Theor.\ Math.\ Phys.\  {\bf 6}, 1135 (2003)
  [arXiv:gr-qc/0212096];
J.~Natario and R.~Schiappa,
{\it ``On the Classification of Asymptotic Quasinormal Frequencies
for $d$--Dimensional Black Holes and Quantum Gravity,''}
Adv. Theor. Math. Phys.  {\bf 8}, 1001 (2004)
[arXiv:hep-th/0411267];   M.~R.~Setare,
  Class.\ Quant.\ Grav.\  {\bf 21}, 1453 (2004)
  [arXiv:hep-th/0311221];
  M.~R.~Setare,
   ``Area spectrum of extremal Reissner-Nordstroem black holes from
  Phys.\ Rev.\ D {\bf 69}, 044016 (2004)
  [arXiv:hep-th/0312061].



\bibitem{Bojowald:2004si}
  M.~Bojowald and R.~Swiderski,
   {\it ``Spherically symmetric quantum horizons,''}
  Phys.\ Rev.\ D {\bf 71}, 081501 (2005)
  [arXiv:gr-qc/0410147].

\bibitem{information}
  D.~Beckman, D.~Gottesman, M.~A.~Nielsen and J.~Preskill,
  Phys.\ Rev.\ A {\bf 64}, 052309 (2001),
 B. Schumacher, M. Westmoreland,
  arXiv.org:quant-ph/0406223.


\bibitem{Fredenhagen:1989kr}
  K.~Fredenhagen and R.~Haag,
   {\it ``On The Derivation Of Hawking Radiation Associated With The Formation Of A
  Black Hole,''}
  Commun.\ Math.\ Phys.\  {\bf 127}, 273 (1990).



\bibitem{Krasnov:1996tb}
  K.~V.~Krasnov,
   {\it ``Counting surface states in the loop quantum gravity,''}
  Phys.\ Rev.\ D {\bf 55}, 3505 (1997)
  [arXiv:gr-qc/9603025],
  K.~V.~Krasnov,
   {\it ``On statistical mechanics of gravitational systems,''}
  Gen.\ Rel.\ Grav.\  {\bf 30}, 53 (1998)
  [arXiv:gr-qc/9605047].

\bibitem{Rovelli:1996dv}
  C.~Rovelli,
  {\it``Black hole entropy from loop quantum gravity,''}
  Phys.\ Rev.\ Lett.\  {\bf 77}, 3288 (1996)
  [arXiv:gr-qc/9603063].




\bibitem{Livine:2005mw}
  E. Livine et. al.
   {\it ``Quantum black holes: Entropy and entanglement on the
  horizon,''}
  [arXiv:gr-qc/0508085];
  E. Livine et. al.
   {\it ``Reconstructing quantum geometry from quantum information: Area
  renormalisation, coarse-graining and entanglement on spin
  networks,''}
  [arXiv:gr-qc/0603008];
  D.~R.~Terno,
   {\it ``From qubits to black holes: Entropy, entanglement and all
  that,''}
  Int.\ J.\ Mod.\ Phys.\ D {\bf 14}, 2307 (2005)
  [arXiv:gr-qc/0505068].

\bibitem{Husain:2005jx}
  V.~Husain and O.~Winkler,
   {\it ``Quantum black holes from null expansion operators,''}
  Class.\ Quant.\ Grav.\  {\bf 22}, L135 (2005),
  V.~Husain and O.~Winkler,
  Phys.\ Rev.\ D {\bf 73}, 124007 (2006)
  [arXiv:gr-qc/0601082],
  V.~Husain and O.~Winkler,
  Class.\ Quant.\ Grav.\  {\bf 22}, L127 (2005)
  [arXiv:gr-qc/0410125].

\bibitem{Dittrich:2005sy}
  B.~Dittrich and R.~Loll,
   {\it ``Counting a black hole in Lorentzian product
  triangulations,''}
  [arXiv:gr-qc/0506035].

\bibitem{amb98}
  J Ambjorn and R Loll, {\it ``Non-perturbative Lorentzian Quantum Gravity, Causality and Topology Change,''} Nucl. Phys. B {\bf 536} (1998) 407,
[hep-th/9805108].

\bibitem{Ansari:2005uz}
M. Ansari and F.~Markopoulou,
   {\it ``A statistical formalism of causal dynamical
  triangulations,''}
  Nucl.\ Phys.\ B {\bf 726}, 494 (2005)
  [arXiv:hep-th/0505165].





\bibitem{Frittelli:1996cj}
  S.~Frittelli, L.~Lehner and C.~Rovelli,
   {\it ``The complete spectrum of the area from recoupling theory in loop  quantum
  gravity,''}
  Class.\ Quant.\ Grav.\  {\bf 13}, 2921 (1996)
  [arXiv:gr-qc/9608043].

\bibitem{Ansari2006locality}
M. Ansari, in preparation.



\bibitem{Ambjorn:2001cv}
  J.~Ambjorn, J.~Jurkiewicz and R.~Loll,
   {\it ``Dynamically triangulating Lorentzian quantum gravity,''}
  Nucl.\ Phys.\ B {\bf 610}, 347 (2001)
  [arXiv:hep-th/0105267].




\bibitem{Sorkin:2005qx}
  R.~D.~Sorkin,
   {\it ``Ten theses on black hole entropy,''}
  Stud.\ Hist.\ Philos.\ Mod.\ Phys.\  {\bf 36}, 291 (2005)
  [arXiv:hep-th/0504037].

\bibitem{Barger}
S. F. Barger,  {\it ``Solution to problem 10454, Amenable
Numbers''}, Amer. Math. Monthly Vol. 105 No. 4 April 1998 MAA
Washington DC.




\bibitem{Gerlach}
U. Gerlach, preceding paper, Phys. Rev. D 14, 1479 (1976).

\bibitem{Cline:2001tq}
  D.~B.~Cline, C.~Matthey and S.~Otwinowski,
  {\it  ``Evidence for a Galactic Origin of Very Short Gamma Ray Bursts and Primordial Black Hole Sources,''}
  Astropart.\ Phys.\  {\bf 18}, 531 (2003)
  [arXiv:astro-ph/0110276];
  D.~B.~Cline, B.~Czerny, C.~Matthey, A.~Janiuk and S.~Otwinowski,
  {\it ``Study of Very Short GRB: New Results from BATSE and
  KONUS,''}
  Astrophys.\ J.\  {\bf 633}, L73 (2005)
  [arXiv:astro-ph/0510309];
  B.~J.~Carr,
  {\it ``Primordial Black Holes: Do They Exist and Are They
  Useful?,''}
  [arXiv:astro-ph/0511743];
   A.~Rau, A.~v.~Kienlin, K.~Hurley and G.~G.~Lichti,
  {\it ``The 1st INTEGRAL SPI-ACS Gamma-Ray Burst Catalogue,''}
  Astron.\ Astrophys.\  {\bf 438}, 1175 (2005)
  [arXiv:astro-ph/0504357].




 \bibitem{Frampton}
  P.~H.~Frampton and T.~W.~Kephart,
  {\it ``Primordial black holes, Hawking radiation and the early
  universe,''}
  Mod.\ Phys.\ Lett.\ A {\bf 20}, 1573 (2005)
  [arXiv:hep-ph/0503267];
    I.~B.~Khriplovich and N.~Produit,
   {\it ``Is Radiation of Black Holes Observable?,''}
  [arXiv:astro-ph/0604003];
  G.~Bertone,
  {\it ``Dark matter: The connection with gamma-ray astrophysics,''}
  [arXiv:astro-ph/0608706];
    P.~Sizun, M.~Casse and S.~Schanne,
   {\it ``Continuum gamma-ray emission from light dark matter positrons and
   electrons,''}
  [arXiv:astro-ph/0607374];
   F.~Ferrer and T.~Vachaspati,
  {\it ``511-keV photons from superconducting cosmic strings,''}
  Phys.\ Rev.\ Lett.\  {\bf 95}, 261302 (2005)
  [arXiv:astro-ph/0505063];
    G.~Bertone, A.~Kusenko, S.~Palomares-Ruiz, S.~Pascoli and D.~Semikoz,
  {\it ``Gamma ray bursts and the origin of galactic positrons,''}
  Phys.\ Lett.\ B {\bf 636}, 20 (2006)
  [arXiv:astro-ph/0405005].


\bibitem{Barcelo:2001ca}
  C.~Barcelo, S.~Liberati and M.~Visser,
   {\it ``Towards the observation of Hawking radiation in Bose-Einstein
  condensates,''}
  Int.\ J.\ Mod.\ Phys.\ A {\bf 18}, 3735 (2003)
  [arXiv:gr-qc/0110036];
S.~Hossenfelder, M.~Bleicher, S.~Hofmann, H.~Stoecker and
A.~V.~Kotwal,
  { \it``Black hole relics in large extra dimensions,''}
  Phys.\ Lett.\ B {\bf 566} (2003) 233
  [arXiv:hep-ph/0302247].



\bibitem{Niven1966}
Niven et. al.,  {\it ``An Introduction to the Theory of Numbers''},
2nd ed., Wiley, NY, 1966, p. 251; S. Ramanujan, {\it
 ``Irregular numbers''}, J. Indian Math. Soc. 5 (1913) 105-106.

\bibitem{Buell}
D. A. Buell,  {\it ``Binary Quadratic Forms''}. Springer-Verlag, NY,
1989; H. Cohen, {\it ``Course in Computational Alg. No. Theory''},
Springer, 1993, p. 514; P. Ribenboim, {\it ``Algebraic Numbers''},
Wiley, NY, 1972, p. 97.





\end{thebibliography}
\end{document}